\documentclass[%
reprint,
superscriptaddress,
showpacs,
amsmath,amssymb,
prc,
floatfix, ]%
{revtex4-1}
\usepackage{color}
\usepackage{graphicx}
\usepackage{dcolumn}
\usepackage{bm}
\usepackage[dvipdfm,bookmarks=true,colorlinks,%
            citecolor=blue,linkcolor=blue, hypertex, %
            breaklinks=true]{hyperref}

\begin{document}


\title{Systematic investigation of the rotational bands in nuclei with $Z \approx 100$
       using a particle-number conserving method based on a cranked shell model}

\author{Zhen-Hua Zhang}
 \affiliation{Key Laboratory of Frontiers in Theoretical Physics,
              Institute of Theoretical Physics, Chinese Academy of Sciences,
              Beijing 100190, China}
\author{Xiao-Tao He}
 \affiliation{College of Material Science and Technology, Nanjing University of
              Aeronautics and Astronautics, Nanjing 210016, China}
\author{Jin-Yan Zeng}
 \affiliation{School of Physics, Peking University, Beijing 100871, China}
\author{En-Guang Zhao}
 \affiliation{Key Laboratory of Frontiers in Theoretical Physics,
              Institute of Theoretical Physics, Chinese Academy of Sciences,
              Beijing 100190, China}
 \affiliation{Center of Theoretical Nuclear Physics, National Laboratory
              of Heavy Ion Accelerator, Lanzhou 730000, China}
 \affiliation{School of Physics, Peking University,
              Beijing 100871, China}
\author{Shan-Gui Zhou}
 \email{sgzhou@itp.ac.cn}
 \affiliation{Key Laboratory of Frontiers in Theoretical Physics,
              Institute of Theoretical Physics, Chinese Academy of Sciences,
              Beijing 100190, China}
 \affiliation{Center of Theoretical Nuclear Physics, National Laboratory
              of Heavy Ion Accelerator, Lanzhou 730000, China}

\date{\today}

\begin{abstract}
The rotational bands in nuclei with $Z \approx 100$ are investigated systematically
by using a cranked shell model (CSM) with the pairing correlations treated by a particle-number
conserving (PNC) method, in which the blocking effects are taken into account exactly.
By fitting the experimental single-particle spectra in these nuclei,
a new set of Nilsson parameters ($\kappa$ and $\mu$) and
deformation parameters ($\varepsilon_2$ and $\varepsilon_4$) are proposed.
The experimental kinematic moments of inertia for the rotational bands  in even-even,
odd-$A$ and odd-odd nuclei, and the bandhead energies of the 1-quasiparticle
bands in odd-$A$ nuclei, are reproduced quite well by the PNC-CSM calculations.
By analyzing the $\omega$-dependence of the occupation probability of each
cranked Nilsson orbital near the Fermi surface and the contributions of valence orbitals
in each major shell to the angular momentum alignment, the upbending
mechanism in this region is understood clearly.
\end{abstract}


\maketitle

\section{Introduction}

Since the importance of shell effects on the stability of superheavy
nuclei (SHN) was illustrated~\cite{Myers1966_NP81-1} and the existence
of an island of stability of SHN was predicted around $Z=114$ and
$N=184$~\cite{Sobiczewski1966_PL22-500, Meldner1967_ArkivF36-593,
Nilsson1968_NPA115-545, *Nilsson1969_PLB28-458, *Nilsson1969_NPA131-1,
Mosel1969_ZPA222-261, *Grumann1969_ZPA228-371},
a lot of efforts have focused on the exploration of SHN.
Great experimental progresses have been made in synthesizing
the superheavy elements (SHE).
Up to now, SHE's with $Z \le 118$ have been synthesized via cold
and hot fusion reactions~\cite{Hofmann2000_RMP72-733, Morita2004_JPSJ73-2593,
Oganessian2007_JPG34-R165, *Oganessian2010_PRL104-142502}.
However, these SHN are all neutron deficient with neutron numbers
less by at least 7 than the predicted next neutron magic number 184.
Therefore, one still can not make a definite conclusion about
the location of the island of stability.

The single particle shell structure is crucial for the location
of the island of stability.
For example, whether the next shell closure of protons appears
at $Z=114$ or 120 is mainly determined by the splitting of
the spin doublets $\pi2f_{5/2,7/2}$.
Experimentally one can not investigate directly the single particle
level structure of SHN with $Z \ge 110$ because the production cross
sections of these nuclei are tiny which makes the spectroscopy experiment impossible at present.
Theoretically different models usually predict different closed shells
beyond $^{208}$Pb; even within the same model there are parameter-dependent
predictions~\cite{Sobiczewski2007_PPNP58-292}.
For examples, the macroscopic-microscopic models or the extended
Thomas-Fermi-Strutinsky integral approach predict that the next shell
closure for the proton is at $Z = 114$~\cite{Cwiok1992_ZPA342-203,
Moeller1995_ADNDT59-185, Mamdouh2001_NPA679-337};
the relativistic mean field models predict $Z = 114$ or 120 to be
the next proton magic number~\cite{Rutz1997_PRC56-238,
*Bender1999_PRC60-034304, *Kruppa2000_PRC61-034313,
Lalazissis1996_NPA608-202,
Zhang2005_NPA753-106, Meng2006_PPNP57-470};
predictions from non-relativistic mean field models with Skyrme
forces are $Z=114$, 120, 124, or 126, depending on the parametrization~\cite{Rutz1997_PRC56-238,
*Bender1999_PRC60-034304, *Kruppa2000_PRC61-034313}.
The shell structure of SHN is important not only for the location of the
island of stability, but also for the study of the synthesis mechanism of
SHN~\cite{Adamian1998_NPA633-409, *Adamian2004_PRC69-011601, *Zubov2009_PPN40-847,
Li2003_EPL64-750, Zhao2008_IJMPE17-1937, Li2010_NPA834-353c, Feng2011_NPR28-1,
Swiatecki2005_PRC71-014602,
Liu2007_PRC76-034604, *Liu2009_PRC80-034601, *Liu2009_SciChinaG52-1482,
Zagrebaev2001_PRC64-034606, *Zagrebaev2005_JPG31-825, *Zagrebaev2008_PRC78-034610,
Bian2009_NPA829-1, *Du2009_SciChinaG52-1489,
Abe2003_PAN66-1057, *Shen2002_PRC66-061602R, *Shen2009_SciChinaG52-1458},
particularly for the survival probability of the excited compound
nuclei~\cite{Adamian2000_PRC62-064303, *Adamian2009_PRC79-054608,
Xia2011_SciChinaPAM54S1-109}.

To learn more about the shell structure of SHN, an indirect way is to study lighter
nuclei in the deformed region with $Z \approx 100$ and $N \approx 152$.
The strongly downsloping orbitals originating from the spherical subshells
and active in the vicinity of the predicted shell closures may come close
to the Fermi surface in these deformed nuclei.
The rotational properties of nuclei in this mass region are
affected strongly by these spherical orbitals.
For examples, the $\pi 1/2^-[521]$ and $\pi 3/2^-[521]$ orbitals are of
particular interest since they stem from the spherical spin doublets $\pi2f_{5/2,7/2}$ orbtials.

Both the in-beam spectroscopy and spectroscopy following the decay
of isomeric states or alpha decays have been used to study nuclei
with $Z \approx 100$~\cite{Leino2004_ARNPS54-175, Ackermann2011_APPB42-577,
Herzberg2004_JPG30-R123, *Herzberg2008_PPNP61-674, Greenlees2011_APPB42-587}.
These nuclei are well deformed; for examples, the quadrupole deformation
parameter $\beta_2 \approx 0.28$ for $^{250}$Fm and
$^{252,254}$No~\cite{Reiter1999_PRL82-509,
Bastin2006_PRC73-024308, Herzberg2001_PRC65-014303}.
Many high-spin rotational bands in even-even nuclei (e.g.,
$^{248,250,252}$Cf~\cite{Takahashi2010_PRC81-057303},
$^{250}$Fm~\cite{Bastin2006_PRC73-024308}
$^{252, 254}$No~\cite{Herzberg2001_PRC65-014303, Reiter1999_PRL82-509})
and odd-$A$ nuclei (e.g.,
$^{247, 249}$Cm and
$^{249}$Cf~\cite{Tandel2010_PRC82-041301R},
$^{253}$No~\cite{Reiter2005_PRL95-032501, Herzberg2009_EPJA42-333},
$^{251}$Md~\cite{Chatillon2007_PRL98-132503},
$^{255}$Lr~\cite{Ketelhut2009_PRL102-212501})
have been established in recent years.
The study of these nuclei is certainly also very interesting in itself.
The rotational spectra in these nuclei can reveal detailed information
on the single-particle configurations, the shell structure,
the stability against rotation, the high-$K$ isomerism, etc..

Theoretically, the deformations, the shell structure, the rotational
properties, and high-$K$ isomeric states have been studied by using
the self-consistent mean field models~\cite{Afanasjev2003_PRC67-024309,
Bender2003_NPA723-354, Delaroche2006_NPA771-103,
Adamian2010_PPN41-1101, *Adamian2011_PRC84-024324},
the macroscopic-microscopic models~\cite{Cwiok1994_NPA573-356,
Muntian1999_PRC60-041302R, Sobiczewski2001_PRC63-034306,
Parkhomenko2004_APPB35-2447, Parkhomenko2005_APPB36-3115,
Adamian2010_PRC81-024320, *Adamian2010_PRC82-054304,
Adamian2010_PPN41-1101, *Adamian2011_PRC84-024324},
the projected shell model~\cite{Sun2008_PRC77-044307, Chen2008_PRC77-061305R,
Al-Khudair2009_PRC79-034320},
the cranked shell models~\cite{Xu2004_PRL92-252501,
He2009_NPA817-45, Zhang2011_PRC83-011304R},
the quasiparticle (qp) phonon model~\cite{Jolos2011_JPG38-115103},
the particle-triaxial-rotor model~\cite{Zhuang2011_CTP},
the heavy shell model~\cite{Cui2011_in-prep},
etc..
Continuing the recently published Rapid Communication~\cite{Zhang2011_PRC83-011304R},
in this work, we present results of a systematic study of the single
particle structure and rotational properties of nuclei with $Z \approx 100$.
The cranked shell model (CSM) with pairing correlations
treated by a particle-number conserving (PNC)
method~\cite{Zeng1983_NPA405-1, Zeng1994_PRC50-1388} is used.
In contrary to the conventional Bardeen-Cooper-Schrieffer (BCS) or 
Hartree-Fock-Bogolyubov (HFB) approach, in the PNC method, the Hamiltonian 
is solved directly in a truncated Fock-space~\cite{Wu1989_PRC39-666}.
So the particle-number is conserved and the Pauli blocking effects
are taken into account exactly.
Noted that the PNC scheme has been implemented both in relativistic 
and nonrelativistic mean field models~\cite{Meng2006_FPC1-38, Pillet2002_NPA697-141}
in which the single-particle states are calculated from self-consistent
mean field potentials instead of the Nilsson potential.

The paper is organized as follows.
A brief introduction of the PNC treatment of pairing correlations within
the CSM and applications of the PNC-CSM are presented in Sec.~\ref{Sec:PNC}.
The numerical details, including a new set of Nilsson parameters
($\kappa, \mu$), the deformation parameters and pairing parameters
are given in Sec.~\ref{Sec:parameters}.
The PNC-CSM calculation results for even-even, odd-$A$ and odd-odd nuclei
are presented in Sec.~\ref{Sec:results}.
A brief summary is given in Sec.~\ref{Sec:summary}.

\section{Theoretical framework}{\label{Sec:PNC}}

The cranked Nilsson Hamiltonian of an axially symmetric nucleus in the
rotating frame reads~\cite{Zeng1994_PRC50-1388, Xin2000_PRC62-067303},
\begin{eqnarray}
 H_\mathrm{CSM}
 & = &
 H_0 + H_\mathrm{P}
 = H_{\rm Nil}-\omega J_x + H_\mathrm{P}
 \ ,
 \label{eq:H_CSM}
\end{eqnarray}
where $H_{\rm Nil}$ is the Nilsson Hamiltonian~\cite{Nilsson1969_NPA131-1},
$-\omega J_x$ is the Coriolis interaction with the cranking frequency
$\omega$ about the $x$ axis (perpendicular to the nuclear symmetry $z$ axis).
$H_{\rm P} = H_{\rm P}(0) + H_{\rm P}(2)$ is the pairing interaction:
\begin{eqnarray}
 H_{\rm P}(0)
 & = &
  -G_{0} \sum_{\xi\eta} a^\dag_{\xi} a^\dag_{\bar{\xi}}
                        a_{\bar{\eta}} a_{\eta}
  \ ,
 \\
 H_{\rm P}(2)
 & = &
  -G_{2} \sum_{\xi\eta} q_{2}(\xi)q_{2}(\eta)
                        a^\dag_{\xi} a^\dag_{\bar{\xi}}
                        a_{\bar{\eta}} a_{\eta}
  \ ,
\end{eqnarray}
where $\bar{\xi}$ ($\bar{\eta}$) labels the time-reversed state of a
Nilsson state $\xi$ ($\eta$), $q_{2}(\xi) = \sqrt{{16\pi}/{5}}
\langle \xi |r^{2}Y_{20} | \xi \rangle$ is the diagonal element of
the stretched quadrupole operator, and $G_0$ and $G_2$ are the
effective strengths of monopole and quadrupole pairing interactions,
respectively.

Instead of the usual single-particle level truncation in common
shell-model calculations, a cranked many-particle configuration
(CMPC) truncation (Fock space truncation) is adopted which is crucial
to make the PNC calculations for low-lying excited states both
workable and sufficiently accurate~\cite{Wu1989_PRC39-666, Molique1997_PRC56-1795}.
An eigenstate of $H_\mathrm{CSM}$ can be written as
\begin{equation}
 | \psi \rangle = \sum_{i} C_i | i \rangle , \qquad (C_i \; \textrm{real}) \ ,
 \label{eq:eigenstate}
\end{equation}
where $| i \rangle$ is a CMPC (an eigenstate of the one-body operator
$H_0$). By diagonalizing $H_\mathrm{CSM}$ in a sufficiently large CMPC space,
sufficiently accurate solutions for low-lying excited eigenstates of
$H_\mathrm{CSM}$ are obtained.

The angular momentum alignment of $|\psi\rangle$ can be separated into
the diagonal and the off-diagonal parts,
\begin{equation}
 \langle \psi | J_x | \psi \rangle =
 \sum_i C_i^2 \langle i | J_x | i \rangle +
 2\sum_{i<j}C_i C_j \langle i | J_x | j \rangle \ ,
 \label{eq:jx}
\end{equation}
and the kinematic moment of inertia (MOI) of $| \psi \rangle$ is
\begin{equation}
 J^{(1)} = \frac{1}{\omega} \langle\psi | J_x | \psi \rangle \ .
\end{equation}
Considering $J_x$ to be a one-body operator, the matrix element
$\langle i | J_x | j \rangle$ for $i\neq j$ is nonzero only when
$|i\rangle$ and $|j\rangle$ differ by one particle
occupation~\cite{Zeng1994_PRC50-1388, Zeng1994_PRC50-746}.
After a certain permutation of creation operators, $|i\rangle$ and
$|j\rangle$ can be recast into
\begin{equation}
 |i\rangle=(-1)^{M_{i\mu}}|\mu\cdots \rangle \ , \qquad
 |j\rangle=(-1)^{M_{j\nu}}|\nu\cdots \rangle \ ,
\end{equation}
where the ellipsis stands for the same particle occupation,
and $(-1)^{M_{i\mu}}=\pm1$, $(-1)^{M_{j\nu}}=\pm1$ according to
whether the permutation is even or odd. Therefore, the kinematic MOI
of $|\psi\rangle$ can be written as
\begin{equation}
 J^{(1)} = \sum_{\mu}j^{(1)}_{\mu}+\sum_{\mu<\nu}j^{(1)}_{\mu\nu} \ ,
\end{equation}
where
\begin{eqnarray}
 j^{(1)}_{\mu}   & = & \frac{n_\mu}{\omega}\langle\mu|j_{x}|\mu\rangle  \ ,
 \nonumber \\
 j^{(1)}_{\mu\nu}& = & \frac{2}{\omega} \langle\mu|j_{x}|\nu\rangle
                       \sum_{i<j} (-1)^{M_{i\mu}+M_{j\nu}} C_{i}C_{j} \ ,
  \qquad (\mu\neq\nu) \ ,
 \nonumber
\end{eqnarray}
and
\begin{equation}
 n_{\mu} = \sum_{i}|C_{i}|^{2}P_{i\mu} \ ,
\end{equation}
is the occupation probability of the cranked orbital $|\mu\rangle$,
$P_{i\mu}=1$ if $|\mu\rangle$ is occupied in $|i\rangle$, and
$P_{i\mu}=0$ otherwise.

We note that because $R_{x}(\pi)=e^{-i \pi J_x}$, $[J_{x}, J_{z}]\neq0$,
the signature scheme breaks the quantum number $K$.
However, it has been pointed out that~\cite{Zeng1994_PRC50-1388, Wu1990_PRC41-1822},
although $[J_{x}, J_{z}] \neq 0$, $[R_{x}(\pi), J_{z}^2] = 0$.
Thus we can construct simultaneous eigenstates of ($R_{x}(\pi), J_{z}^2$).
Each CMPC $|i\rangle$ in Eq.~(\ref{eq:eigenstate}) is chosen as a
simultaneous eigenstate of ($H_0, J_{z}^2$).
It should be noted that, though the projection $K$ of
the total angular momentum of a deformed spheroidal nucleus is a
constant of motion, $K$ can not keep constant when the rotational frequency
$\omega$ is non-zero due to the Coriolis interaction.
However, in the low-$\omega$ region, $K$ may be served as
an useful quantum number characterizing a low-lying
excited rotational band.

The PNC-CSM treatment has been used to describe successfully the high-spin
states of the normally deformed nuclei in the rare-earth and
the actinide region, and the superdeformed nuclei in $A \approx 190$ region.
The multi-qp high-$K$ isomer states are investigated in detail
in the well-deformed Lu ($Z$=71), Hf ($Z$=72) and Ta ($Z$=73)
isotopes~\cite{Zhang2009_NPA816-19, Zhang2009_PRC80-034313}.
The backbendings in Yb ($Z$=70) and Tm ($Z$=68) isotopes are understood
clearly, especially the occurrence of sharp backbending in some
nuclei~\cite{Liu2004_NPA735-77, Chen2009_SciChinaG52-1542}.
The upbending mechanisms in the actinide nuclei $^{251}$Md and
$^{253}$No are also analyzed~\cite{He2009_NPA817-45}.
In the superdeformed nuclei around $A \approx 190$ region, the effects
of the quadrupole pairing on the downturn of dynamic MOI's are
analyzed~\cite{Xin2000_PRC62-067303}, the evolution of the
dynamic MOI's and the alignment additivity, which come from the
contributions of the interference terms, have been
investigated~\cite{He2005_NPA760-263}.

Some general features in nuclear structure physics have also been
explained well in the PNC-CSM scheme, e.g.,
the large fluctuations of odd-even differences in
MOI's~\cite{Zeng1994_PRC50-746}, the nonadditivity in
MOI's~\cite{Liu2002_PRC66-067301}, the microscopic mechanism of
identical bands in normally deformed nuclei
and super-deformed nuclei~\cite{Zeng2001_PRC63-024305, Liu2002_PRC66-024320},
the non-existence of the pairing phase transition~\cite{Wu2011_PRC83-034323}, etc..

In Ref.~\cite{Zhang2011_PRC83-011304R}, the high-spin rotational bands
in $^{247, 249}$Cm and $^{249}$Cf established in Ref.~\cite{Tandel2010_PRC82-041301R}
have been calculated using the PNC-CSM method and the upbending mechanism are discussed.
This paper is an extension of \cite{Zhang2011_PRC83-011304R}, providing more details
and a systematic investigation of nuclei with $Z \approx 100$.

\section{Numerical details}{\label{Sec:parameters}}

\subsection{A new set of Nilsson parameters}
\begin{table}[h]
\caption{\label{tab:ku}
Nilsson parameters $\kappa$ and $\mu$ proposed for the nuclei with $Z \approx$
100, which has been given in Ref.~\protect\cite{Zhang2011_PRC83-011304R}.
}
\begin{center}
\def\temptablewidth{8.6cm}
\begin{tabular*}
{\temptablewidth}{@{\extracolsep{\fill}}cccc|cccc}
\hline\hline
 $N$ & $l$     & $\kappa_p$ & $\mu_p$ &
 $N$ & $l$     & $\kappa_n$ & $\mu_n$ \\
\hline
  4  & 0,2,4   & 0.0670     & 0.654   &
     &         &            &         \\
  5  & 1       & 0.0250     & 0.710   &
  6  & 0       & 0.1600     & 0.320   \\
     & 3       & 0.0570     & 0.800   &
     & 2       & 0.0640     & 0.200   \\
     & 5       & 0.0570     & 0.710   &
     & 4,6     & 0.0680     & 0.260   \\
  6  & 0,2,4,6 & 0.0570     & 0.654   &
  7  & 1,3,5,7 & 0.0634     & 0.318   \\
\hline\hline
\end{tabular*}
\end{center}
\end{table}

The conventional Nilsson parameters ($\kappa, \mu$) proposed in
Refs.~\cite{Nilsson1969_NPA131-1, Bengtsson1985_NPA436-14} are optimized to
reproduce the experimental level schemes for the rare-earth and actinide nuclei.
However, these two sets of parameters can not describe well the experimental
level schemes of nuclei studied in this work.
By fitting the experimental single-particle levels in the odd-$A$ nuclei
with $Z \approx 100$, we obtained a new set of Nilsson parameters ($\kappa$, $\mu$)
which are dependent on the main oscillator quantum number $N$ as well as
the orbital angular momentum $l$~\cite{Zhang2011_PRC83-011304R}.
Here for the completeness, we include them again in Table~\ref{tab:ku}.
Note that the readjustment of Nilsson parameters is also necessary in
some other mass regions of the nuclear chart~\cite{Seo1986_ZPA324-43,
Zhang1998_PRC58-2663R, Sun2000_PRC62-021601}
and the $l$-dependence was already included in Refs.~\cite{Seo1986_ZPA324-43,
Zhang1998_PRC58-2663R}.

\subsection{Deformation parameters}

\begin{figure}
\includegraphics[scale=0.38]{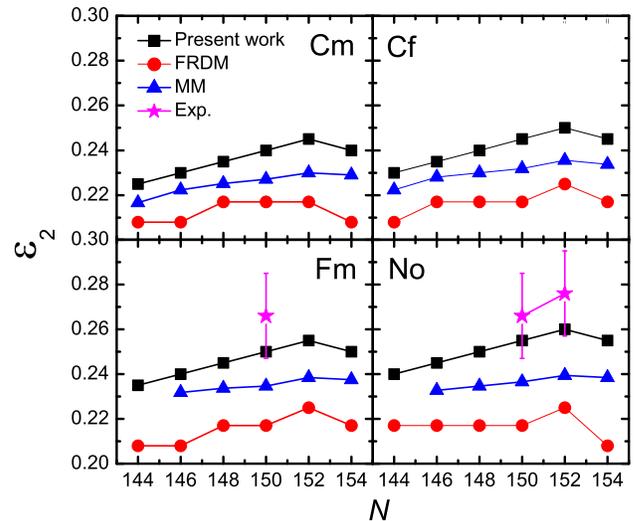}
\caption{\label{fig:Deformation}(Color online)
The quadrupole deformations given in Table~\ref{tab:deformation} (black squares)
and those predicted by a macroscopic-microscopic (MM)
model~\cite{Cwiok1994_NPA573-356, Sobiczewski2001_PRC63-034306} (blue triangles)
and the finite range droplet model (FRDM)~\cite{Moeller1995_ADNDT59-185}
(red solid circles).
The experimental values for $^{250}$Fm~\cite{Bastin2006_PRC73-024308}
and $^{252,254}$No~\cite{Herzberg2001_PRC65-014303} (pink stars) are
also shown for comparison.
}
\end{figure}

\begin{table}
\caption{\label{tab:deformation}%
Deformation parameters $\varepsilon_2$ and $\varepsilon_4$ used in the PNC-CSM
calculations for even-even nuclei with $Z \approx 100$.
}
\begin{center}
\def\temptablewidth{8.5cm}
\begin{tabular*}{\temptablewidth}{@{\extracolsep{\fill}}ccccccc}
\hline\hline
 $N$            &$  144$    &$  146$    &$  148$    &$  150$    &$  152$    &$  154$    \\
\hline
 $Z=96$         &$^{240}$Cm &$^{242}$Cm &$^{244}$Cm &$^{246}$Cm &$^{248}$Cm &$^{250}$Cm \\
$\varepsilon_2$ &    0.225  &    0.230  &    0.235  &    0.240  &    0.245  &    0.240  \\
$\varepsilon_4$ & $-$0.015  & $-$0.010  & $-$0.005  &    0.000  &    0.005  &    0.01   \\
\hline
 $Z=98$         &$^{242}$Cf &$^{244}$Cf &$^{246}$Cf &$^{248}$Cf &$^{250}$Cf &$^{252}$Cf \\
$\varepsilon_2$ &    0.230  &    0.235  &    0.240  &    0.245  &    0.250  &    0.245  \\
$\varepsilon_4$ & $-$0.01   & $-$0.005  &    0.000  &    0.005  &    0.010  &    0.015  \\
\hline
 $Z=100$        &$^{244}$Fm &$^{246}$Fm &$^{248}$Fm &$^{250}$Fm &$^{252}$Fm &$^{254}$Fm \\
$\varepsilon_2$ &    0.235  &    0.240  &    0.245  &    0.250  &    0.255  &    0.250  \\
$\varepsilon_4$ & $-$0.005  &    0.000  &    0.005  &    0.010  &    0.015  &    0.02   \\
\hline
 $Z=102$        &$^{246}$No &$^{248}$No &$^{250}$No &$^{252}$No &$^{254}$No &$^{256}$No \\
$\varepsilon_2$ &    0.240  &    0.245  &    0.250  &    0.255  &    0.260  &    0.255  \\
$\varepsilon_4$ &    0      &    0.005  &    0.010  &    0.015  &    0.020  &    0.025  \\
\hline\hline
\end{tabular*}
\end{center}
\end{table}

There are not enough experimental values of the deformation
parameters for the very heavy nuclei with $Z \approx$ 100.
According to the data, the quadrupole deformation parameter
$\beta_2 \approx 0.28 \pm 0.02$ for ${}^{250}$Fm~\cite{Bastin2006_PRC73-024308}, ${}^{252,254}$No~\cite{Herzberg2001_PRC65-014303}.
The values used in various calculations or predicted by different models are quite different.
Figure.~\ref{fig:Deformation} shows the experimental quadrupole
deformation (pink stars in Fig.~\ref{fig:Deformation}) and those predicted in the
macroscopic-microscopic (MM) model~\cite{Cwiok1994_NPA573-356, Sobiczewski2001_PRC63-034306}
(blue triangles in Fig.~\ref{fig:Deformation}) and the finite range droplet model
(FRDM)~\cite{Moeller1995_ADNDT59-185} (red solid circles in Fig.~\ref{fig:Deformation}).
The deformation parameters given in these two MM models~\cite{Cwiok1994_NPA573-356,
Sobiczewski2001_PRC63-034306} are very close to each other, hence we only
show the results given in Ref.~\cite{Sobiczewski2001_PRC63-034306}.
From Fig.~\ref{fig:Deformation} we can see that the deformation parameters from MM and FRDM do not
agree with the experimental values, especially those from the FRDM.
One can find a general trend in Fig.~\ref{fig:Deformation}, i.e.,
both the experimental values and the predicted values indicate that the
deformations reach maximum at ${}^{254}$No ($Z=102$ and $N=152$)
partly according to which we fix the deformation parameters used in the present study.

The deformations are input parameters in the PNC-CSM calculations
(black squares in Fig.~\ref{fig:Deformation}).
They are chosen to be close to existed
experimental values and change smoothly according to the proton and the neutron number.
The deformation parameters $\varepsilon_2$ and $\varepsilon_4$ used in
our PNC-CSM calculations for even-even nuclei with $Z\approx 100$
are listed in Table~\ref{tab:deformation}.
The deformations of odd-$A$ and odd-odd nuclei are taken as the average
of the neighboring even-even nuclei.
These parameters we choose may have some discrepancy from the empirical values which may
lead to some deviations in the single-particle levels,
if the single particle levels are very close to each other.
For example, the level sequence of the 1-qp bands will change in an
isotonic or an isotropic chain (e.g., see Figs.~\ref{fig:SPLN147} and \ref{fig:SPLEs})
due to the deformation staggering in the neighboring nuclei.
From Fig.~\ref{fig:Deformation} it can be seen
that this staggering existed in the FRDM.
Some previous calculations have shown that the octupole correlations play important 
roles in this mass 
region~\cite{Shneidman2006_PRC74-034316, Chen2008_PRC77-061305R, Lu2011_arxiv1110.6769v1}.
In the present work, the octupole deformation is not included yet. 
We note that the octupole effect may modify the single-particle level scheme. 
One example will be discussed later for the low excitation energy of the $\nu5/2^+[622]$ 
states in the $N = 151$ isotones in Sec.~\ref{sec:1-qp}.

\subsection{Pairing strengthes and the CMPC space}

\begin{figure}[h]
\includegraphics[scale=0.3]{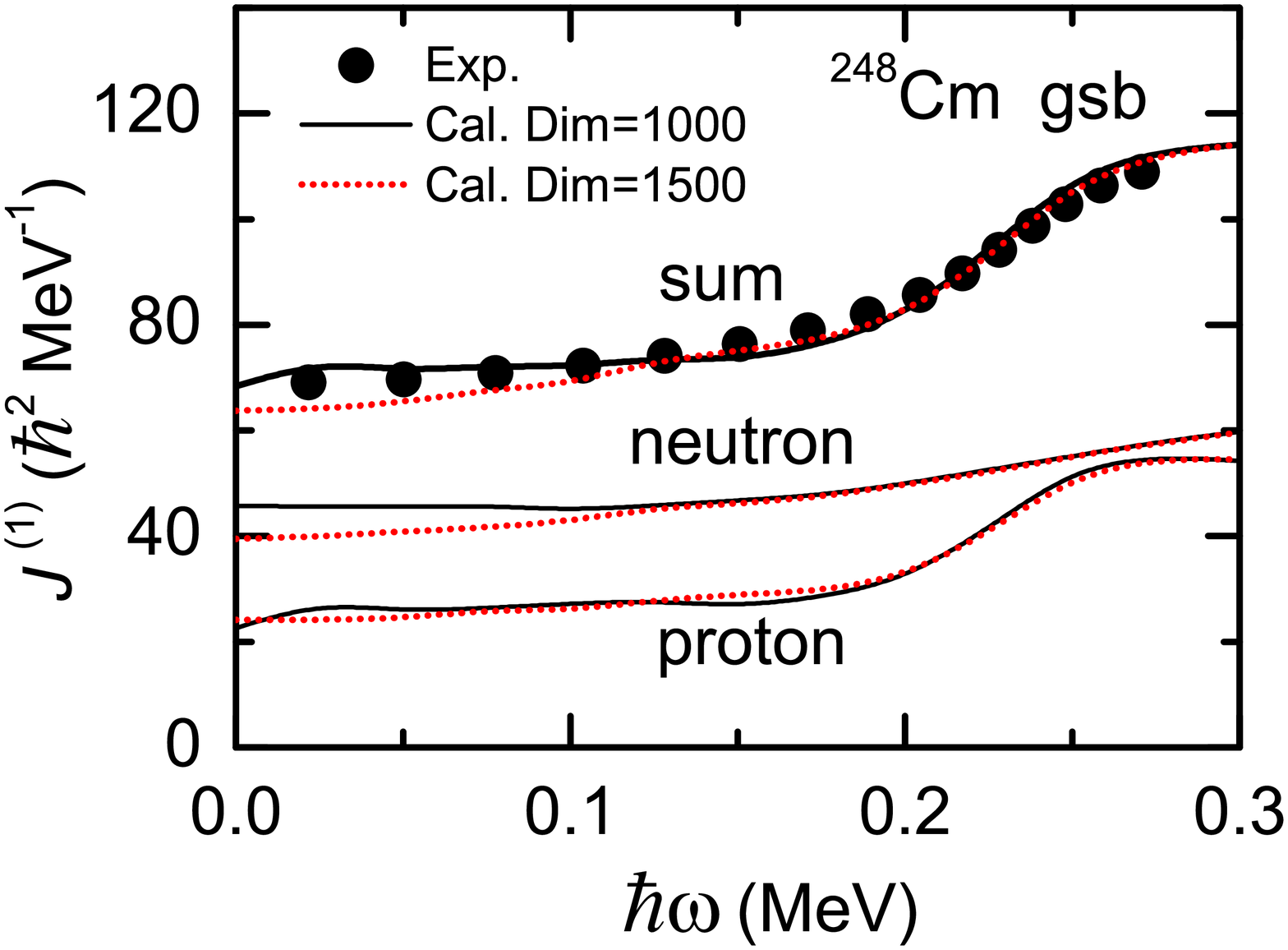}
\caption{\label{fig:248Cm}(Color online)
The calculated MOI's of the GSB in $^{248}$Cm
using different dimensions of the CMPC space [1000 (black solid lines) and 1500
(red dotted lines), respectively]. The experimental values are denoted by solid
circles. The effective pairing strengths used in the calculation, when the
dimensions are 1500 (1000), are $G_p$ = 0.35~MeV, $G_{2p}$ = 0.03~MeV,
$G_n$ = 0.27~MeV, $G_{2n}$ = 0.013~MeV ($G_p$ = 0.40~MeV, $G_{2p}$ = 0.035~MeV,
$G_n$ = 0.30~MeV, $G_{2n}$ = 0.020~MeV as given in Table~\ref{tab:Pairing}).
}
\end{figure}

The effective pairing strengths $G_0$ and $G_2$ can be determined
by the odd-even differences in nuclear binding energies.
They are connected with the dimension of the truncated CMPC space.
The CMPC space for the heavy nuclei studied in this work is constructed
in the proton $N=4, 5, 6$ shells and the neutron $N=6, 7$ shells.
The dimensions of the CMPC space for the nuclei with $Z \approx 100$
are about 1000 both for protons and neutrons.
The corresponding effective monopole and quadrupole pairing strengths are
shown in Table~\ref{tab:Pairing}.
As we are only interested in the yrast and low-lying excited states,
the number of the important CMPC's involved (weight $>1\%$) is not very large
(usually $<20$) and almost all the CMPC's with weight $>0.1\%$ are included in.

The stability of the PNC calculation results against the change of
the dimension of the CMPC space has been investigated in
Refs.~\cite{Zeng1994_PRC50-1388, Liu2002_PRC66-024320, Molique1997_PRC56-1795}.
A larger CMPC space with renormalized pairing strengths gives essentially
the same results.
The calculated MOI's of the ground state band (GSB) in $^{248}$Cm using different
dimensions of the CMPC space (1000 and 1500, respectively) are compared in Fig.~\ref{fig:248Cm}.
The red dotted line is the result calculated with the dimensions of CMPC space
about 1500 both for protons and neutrons.
The effective pairing strengths used in the calculation are $G_p$ = 0.35~MeV,
$G_{2p}$ = 0.03~MeV, $G_n$ = 0.27~MeV, and $G_{2n}$ = 0.013~MeV,
which are a little smaller than those used when the CMPC
dimensions are about 1000, i.e., $G_p$ = 0.40~MeV, $G_{2p}$ = 0.035~MeV,
$G_n$ = 0.30~MeV, and $G_{2n}$ = 0.020~MeV (see Table~\ref{tab:Pairing}).
We can see that these two results agree well with each other and
they both also agree well with the experiment.
So the solutions to the low-lying excited states are quite satisfactory.

\begin{table}[!h]
\caption{\label{tab:Pairing}%
Effective pairing strengths used in the PNC-CSM calculations for the nuclei with $Z \approx$ 100.
}
\begin{center}
\def\temptablewidth{8.6cm}
\begin{tabular*}
{\temptablewidth}{@{\extracolsep{\fill}}ccccc}
\hline\hline
              &even-even & odd-$N$ & odd-$Z$ & odd-odd \\
\hline
$G_{0p}$ (MeV)& 0.40     & 0.40    & 0.25    & 0.25    \\
$G_{2p}$ (MeV)& 0.035    & 0.035   & 0.010   & 0.010   \\
$G_{0n}$ (MeV)& 0.30     & 0.25    & 0.30    & 0.25    \\
$G_{2n}$ (MeV)& 0.020    & 0.015   & 0.020   & 0.015   \\
\hline\hline
\end{tabular*}
\end{center}
\end{table}

\section{Results and discussions}{\label{Sec:results}}

\subsection{Cranked Nilsson levels}

\begin{figure*}
\includegraphics[scale=0.5]{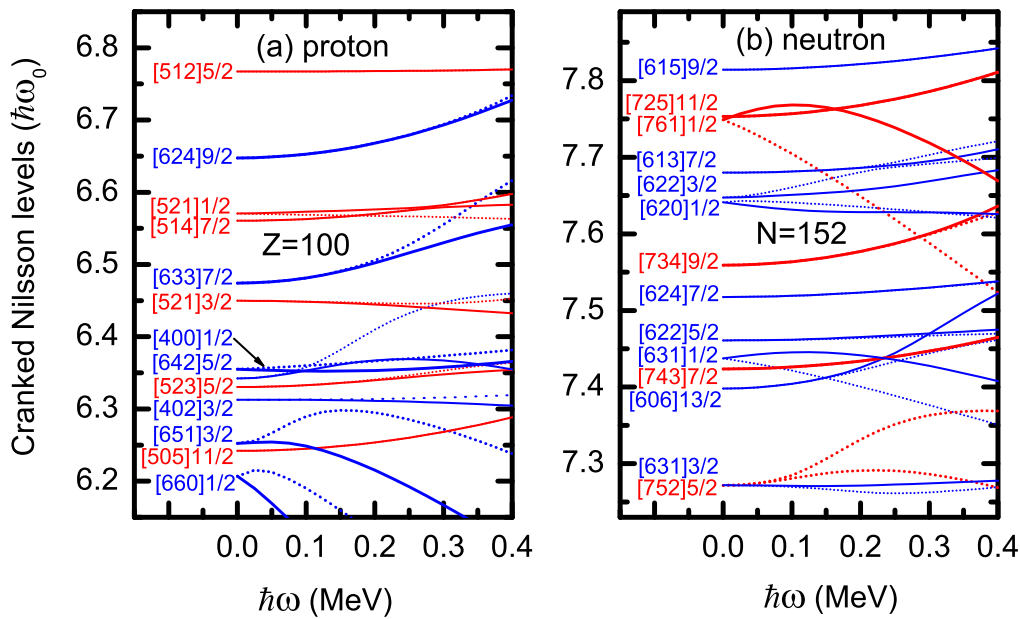}
\caption{\label{fig:250FmNilsson}(Color online)
The cranked Nilsson levels near the Fermi surface of $^{250}$Fm
(a) for protons and (b) for neutrons. The positive (negative) parity
levels are denoted by blue (red) lines. The signature $\alpha=+1/2$
($\alpha=-1/2$) levels are denoted by solid (dotted) lines.
}
\end{figure*}

Figure~\ref{fig:250FmNilsson} shows the calculated cranked Nilsson levels
near the Fermi surface of $^{250}$Fm.
The positive (negative) parity levels are denoted by blue (red) lines.
The signature $\alpha=+1/2$ ($\alpha=-1/2$) levels are denoted by solid (dotted) lines.
For protons, the sequence of single-particle levels near the Fermi surface
is exactly the same as that determined from the experimental information of
$^{249}$Es~\cite{Herzberg2008_PPNP61-674}.
The sequence of single-neutron levels near the Fermi surface is also
very consistent with the one determined from the experimental information of
$^{251}$Fm~\cite{Herzberg2008_PPNP61-674}, with the only exception of
the $\nu5/2^+[622]$ orbital which will be discussed in next subsection.

From Fig.~\ref{fig:250FmNilsson} it is seen that there exist a proton gap at
$Z=100$ and a neutron gap at $N=152$, which is consistent with the
experiment and the calculation by using a Woods-Saxon
potential~\cite{Chasman1977_RMP49-833, *Chasman1978_RMP50-173}.
The position of the proton orbitals $1/2^-[521]$ and $3/2^-[521]$ is very important,
because they stem from the spherical spin partners $2f_{5/2, 7/2}$ orbitals.
The magnitude of the spin-orbital splitting of these spin partners
determines whether the next proton magic number is $Z=114$ or 120.
Noted that the cranked relativistic Hartree-Bogoliubov theory has been used to investigate
the spin-orbital splitting of this spin partners in Ref.~\cite{Afanasjev2003_PRC67-024309}.
The calculated proton levels indicate that the $Z=120$ gap is large whereas the
$Z=114$ gap is small.

The neutron $1/2^+[620]$ orbital is the highest-lying neutron orbital
from above the $N = 164$ spherical subshell ($2g_{7/2}$), based on which
a high-spin rotational band has been established in
$^{249}$Cm~\cite{Tandel2010_PRC82-041301R}.
This orbital also brings new information toward studying those states
close to the next closed shell for neutrons.

\subsection{\label{sec:1-qp}1-qp bandhead energies}

\begin{figure*}
\includegraphics[scale=0.3]{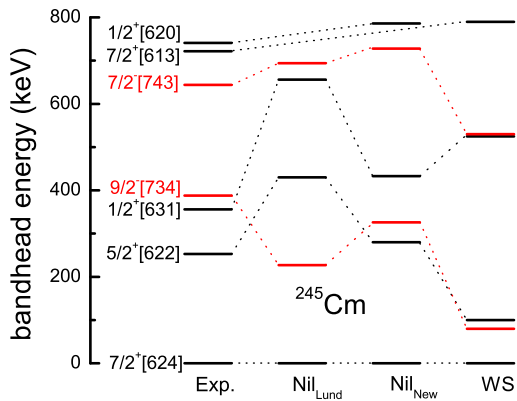}
\caption{\label{fig:CompSPL}(Color online)
Comparison between experimental and theoretical 1-qp bandhead energies for $^{245}$Cm.
The Lund systematic Nilsson parameters are taken from~\cite{Nilsson1969_NPA131-1}.
The results using Woods-Saxon potential are taken from~\cite{Parkhomenko2005_APPB36-3115}.
The positive and negative parity states are denoted by black and red lines, respectively.}
\end{figure*}

First we choose $^{245}$Cm as an example to show the comparison of 
theoretical 1-qp bandhead energies with the data in Fig.~\ref{fig:CompSPL}. 
It can be seen that the calculated results obtained by using the conventional 
Nilsson parameters~\cite{Nilsson1969_NPA131-1} deviate from the experimental 
values very much. With the modified parameters given in Table~\ref{tab:ku}, 
a remarkably good agreement with the data can be achieved. 
For comparison, the results from the Woods-Saxon (WS) 
potential~\cite{Parkhomenko2005_APPB36-3115} is also given 
(the bandhead energies are taken from Table II of 
Ref.~\cite{Parkhomenko2005_APPB36-3115} except that that of $7/2^+[613]$ 
is taken from Fig. 1 of the same article). 
It should be noted that generally speaking, for the 1-qp spectra of nuclei 
with $Z \approx 100$ the Woods-Saxon potential gives a better agreement with the data. 
For example, the root-mean-square deviation of theoretical 1-qp 
bandhead energies from the experimental values is about 270 keV for neutrons 
by using the modified Nilsson parameters, while this deviation is 
about 200 keV by using the Woods-Saxon 
potential~\cite{Parkhomenko2005_APPB36-3115}.

\begin{figure*}
\includegraphics[scale=0.5]{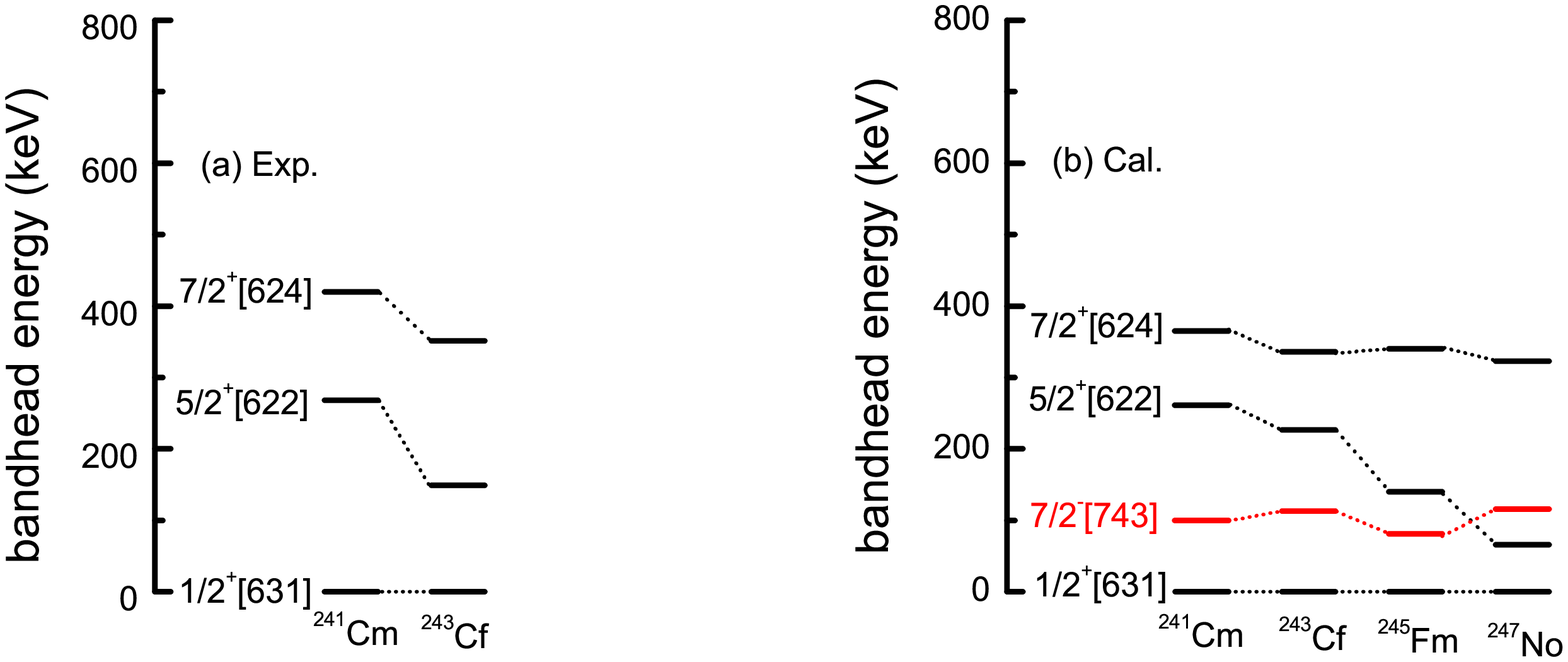}
\caption{\label{fig:SPLN145}(Color online)
(a) Experimental and (b) calculated bandhead energies of low-lying
one-quasineutron bands for the $N=145$ isotones. The data are taken
from~\cite{Martin2005_NDS106-89, Hessberger2004_EPJA22-417}.
The positive and negative parity states are denoted by black and
red lines, respectively.}
\end{figure*}

\begin{figure*}
\includegraphics[scale=0.5]{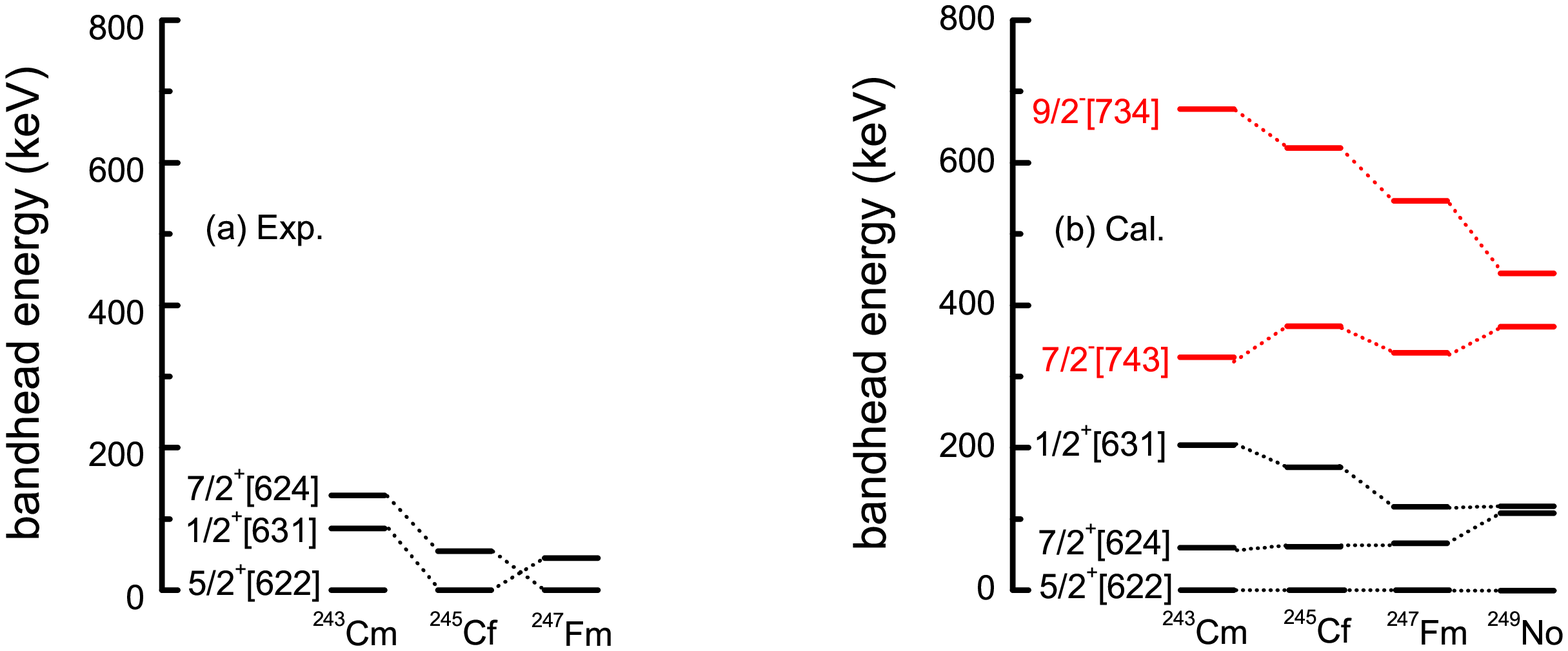}
\caption{\label{fig:SPLN147}(Color online)
(a) Experimental and (b) calculated bandhead energies of low-lying
one-quasineutron bands for the $N=147$ isotones. The data are taken
from~\cite{Braid1971_PRC4-247, Hessberger2004_EPJA22-417, Hessberger2006_EPJA30-561}.
The positive and negative parity states are denoted by black and
red lines, respectively.}
\end{figure*}

\begin{figure*}
\includegraphics[scale=0.5]{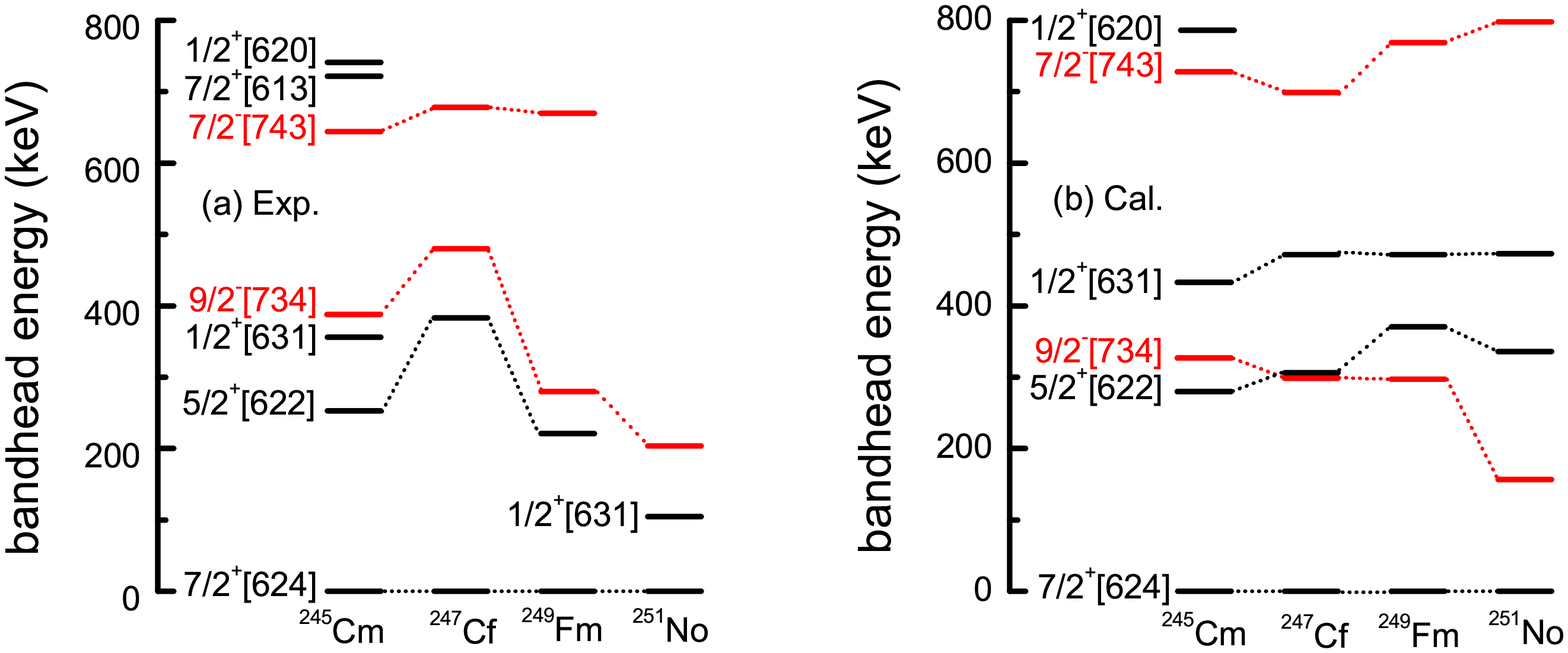}
\caption{\label{fig:SPLN149}(Color online) (a) Experimental and (b) calculated bandhead
energies of low-lying one-quasineutron bands for the $N=149$ isotones.
The data are taken from~\cite{Firestone1999, Akovali2004_NDS102-515,
Hessberger2007_EPJD45-33, Hessberger2006_EPJA30-561}.
The positive and negative parity states are denoted by black and
red lines, respectively.}
\end{figure*}

\begin{figure*}
\includegraphics[scale=0.5]{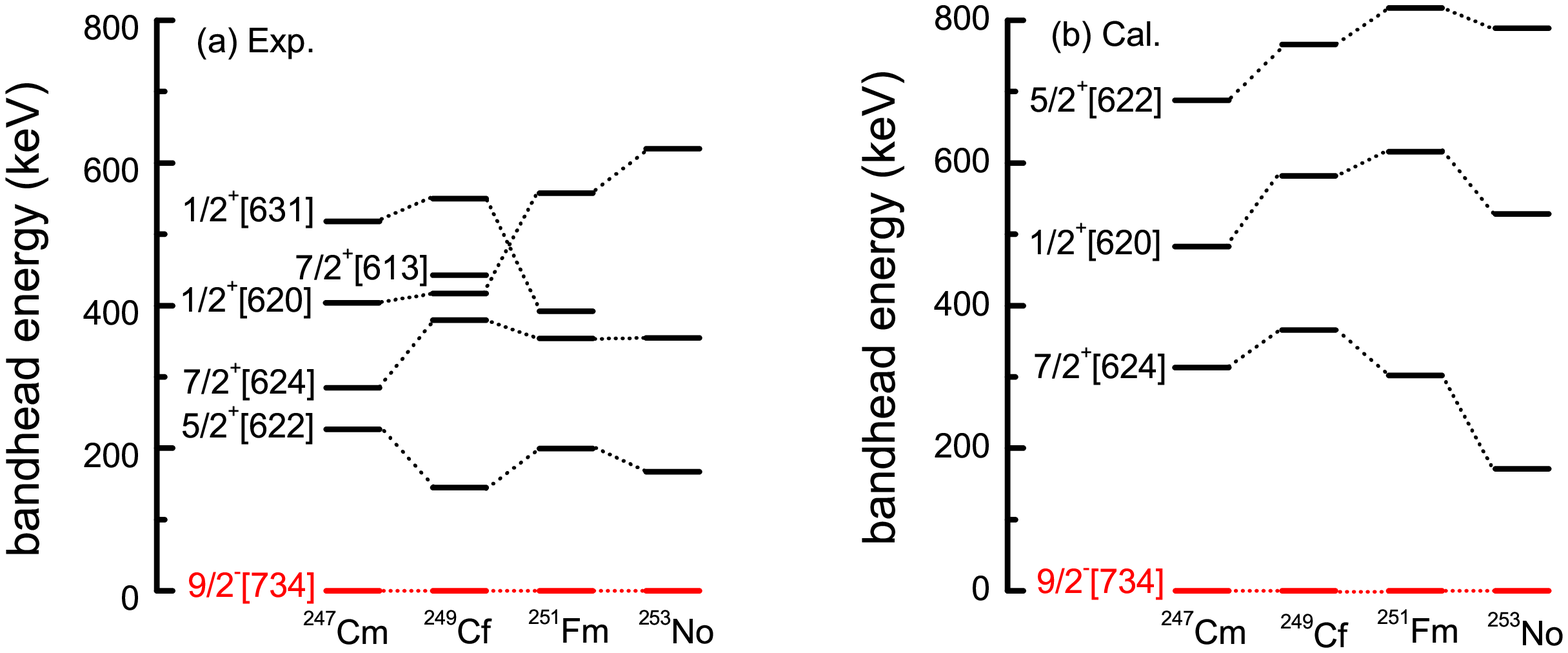}
\caption{\label{fig:SPLN151}(Color online)  (a) Experimental and (b) calculated bandhead
energies of low-lying one-quasineutron bands for the $N=151$ isotones.
The data are taken from~\cite{Ahmad2003_PRC68-044306,
Makii2007_PRC76-061301R, Asai2011_PRC83-014315, Lopez-Martens2007_EPJA32-245}.
The configuration assignment $7/2^+[624]$ in $^{253}$No are taken from
Ref.~\cite{Reiter2005_PRL95-032501} and discussed in Sec.~\ref{Sec:results}.
The positive and negative parity states are denoted by black and
red lines, respectively.}
\end{figure*}

\begin{figure*}
\includegraphics[scale=0.5]{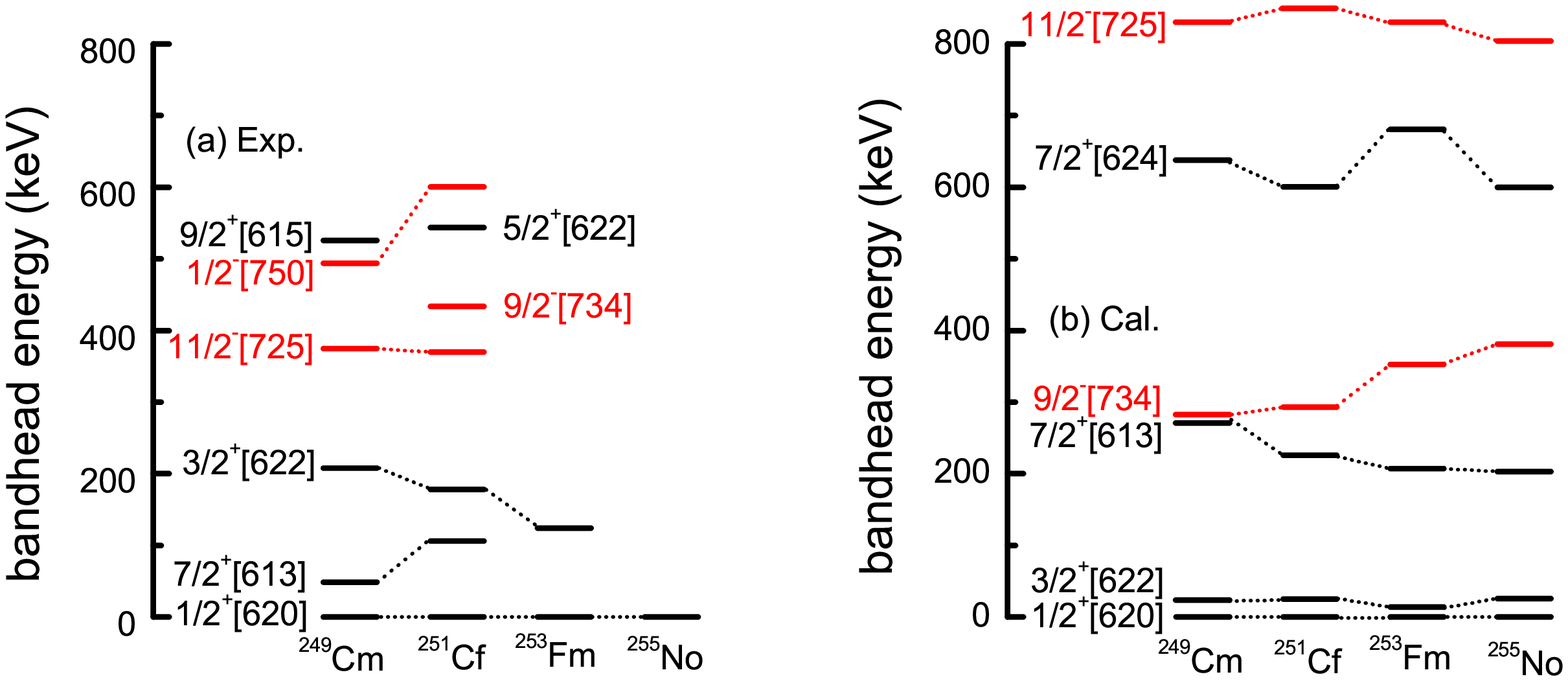}
\caption{\label{fig:SPLN153}(Color online)  (a) Experimental and (b) calculated
bandhead energies of low-lying one-quasineutron bands for the $N=153$ isotones.
The data are taken from~\cite{Ishii2008_PRC78-054309, Ahmad2005_PRC72-054308,
Asai2005_PRL95-102502, Hessberger2006_EPJA29-165}.
The positive and negative parity states are denoted by black and
red lines, respectively.}
\end{figure*}

\begin{figure*}
\includegraphics[scale=0.5]{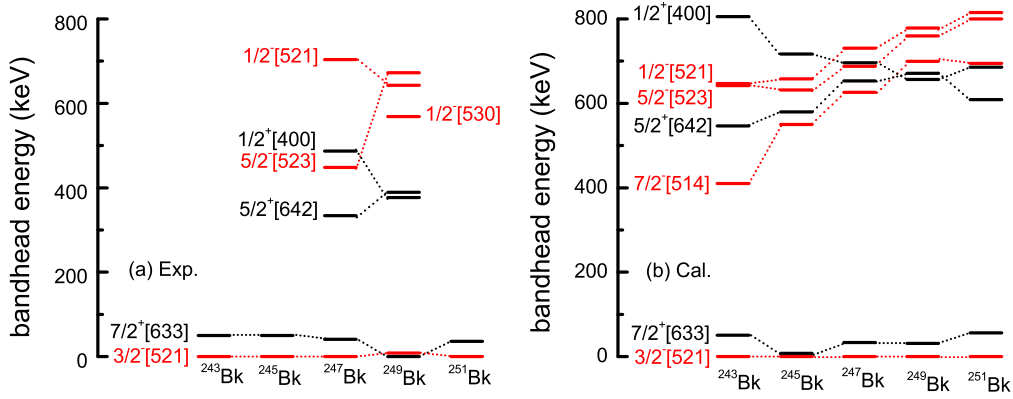}
\caption{\label{fig:SPLBk}(Color online)  (a) Experimental and (b) calculated bandhead
energies of low-lying one-quasiproton bands for the Bk ($Z=97$) isotopes.
The data are taken from~\cite{Yuichi1989_NPA500-90, Ahmad1979_PRC20-290,
Ahmad2005_PRC71-054305, Tuli2006_NDS107-1347}.
The positive and negative parity bands are denoted by black and red lines, respectively.}
\end{figure*}

\begin{figure*}
\includegraphics[scale=0.5]{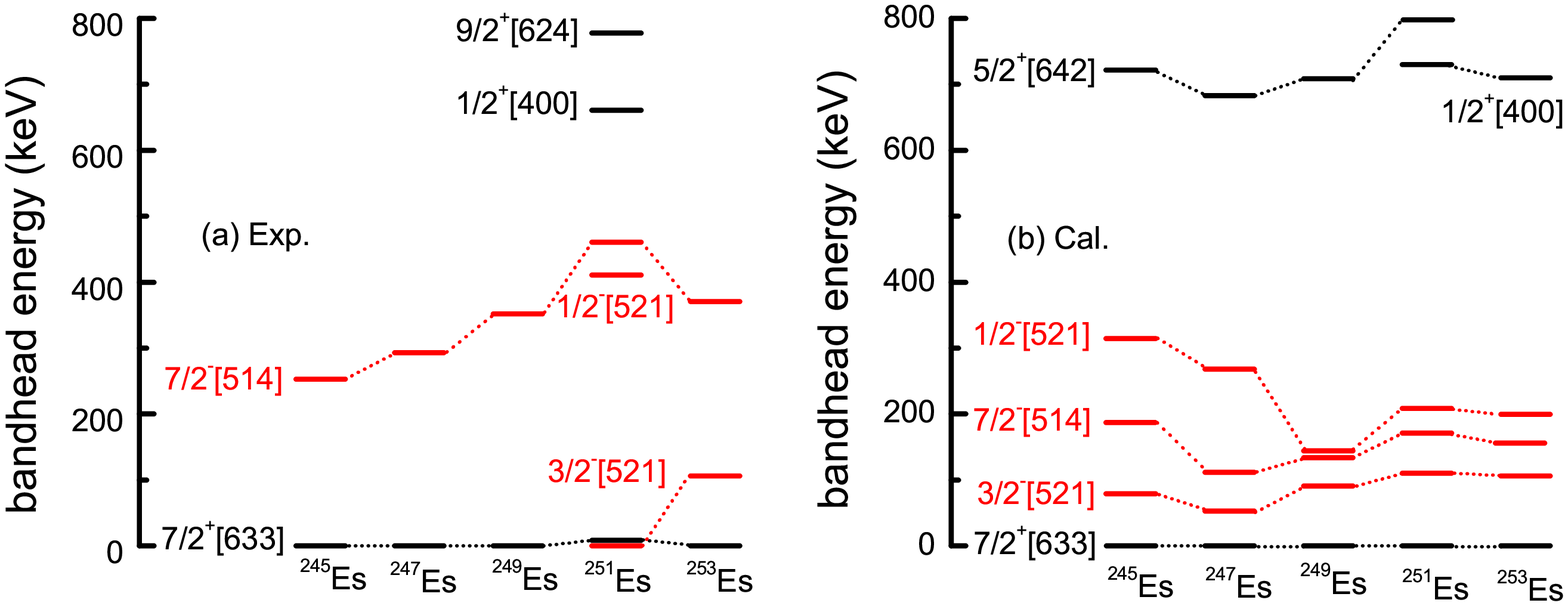}
\caption{\label{fig:SPLEs}(Color online) (a) Experimental and (b) calculated bandhead
energies of low-lying one-quasiproton bands for the Es ($Z=99$) isotopes.
The data are taken from~\cite{Ahmad1978_PRC17-2163, Hessberger2005_EPJA26-233}.
The positive and negative parity states are denoted by black and red lines, respectively.}
\end{figure*}

\begin{figure*}
\includegraphics[scale=0.5]{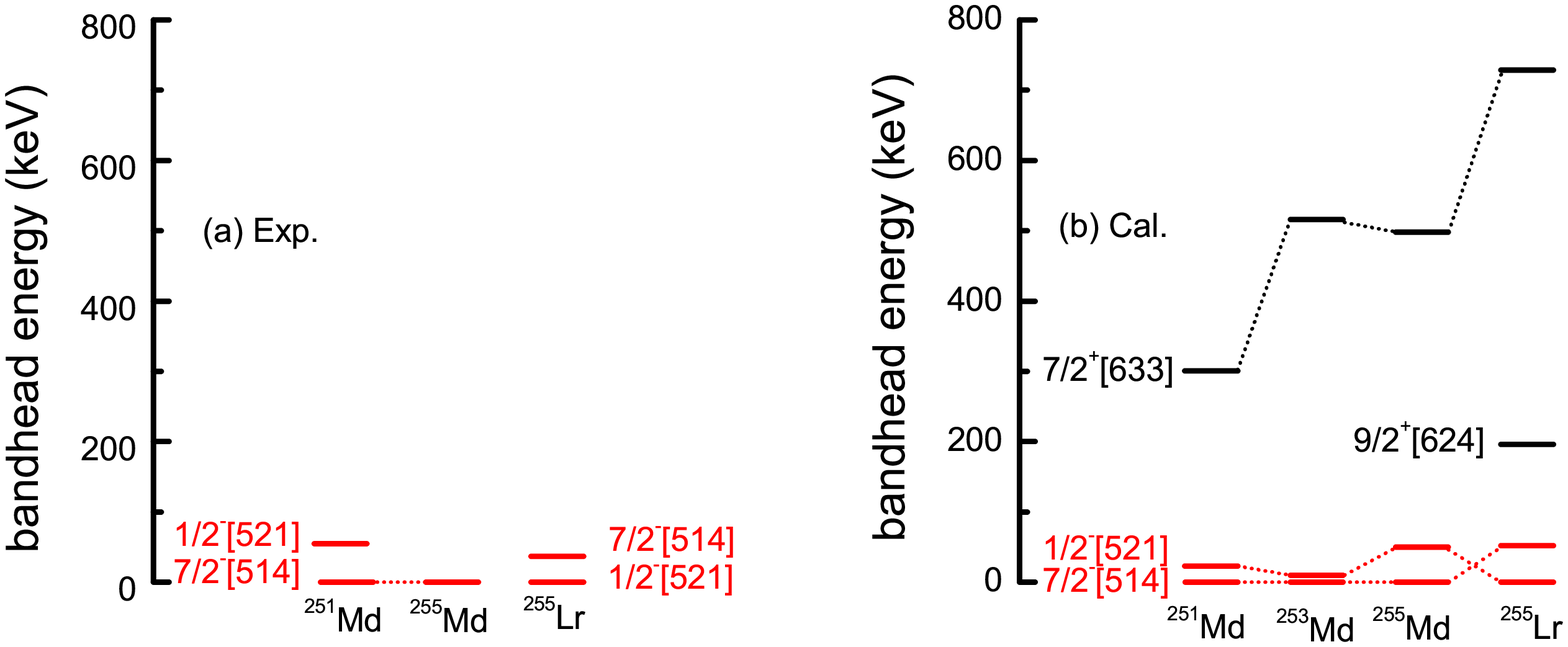}
\caption{\label{fig:SPLMdLr}(Color online) (a) Experimental and (b) calculated bandhead
energies of low-lying one-quasiproton bands for the Md ($Z=101$) and Lr ($Z= 103$) isotopes.
The data are taken from~\cite{Ahmad2000_PRC61-044301, Chatillon2006_EPJA30-397}.
The positive and negative parity states are denoted by black and red lines, respectively.}
\end{figure*}

Figures~\ref{fig:SPLN145}-\ref{fig:SPLMdLr} show experimental and calculated
bandhead energies of the low-lying 1-qp bands for even-$Z$ and
odd-$N$ isotones with $N=145-153$ (one-quasineutron states) and odd-$Z$ and
even-$N$ isotopes with $Z=97-103$ (one-quasiproton states).
In these figures, 1-qp states with energies around or less than
0.8~MeV are shown and the positive and negative parity states are denoted
by black and red lines, respectively.
Generally speaking, the agreement between the calculation and the experiment
is satisfactory.
Next we discuss these 1-qp states in details.

The experimental and calculated bandhead energies of low-lying one-quasineutron
bands for the $N=145$ isotones are compared in Fig.~\ref{fig:SPLN145}.
The data are available only for $^{241}$Cm~\cite{Martin2005_NDS106-89} and
$^{243}$Cf~\cite{Hessberger2004_EPJA22-417} and they are reproduced by
the theory quite well.
The ground state of each $N=145$ isotone studied in this work is $\nu1/2^+[631]$.
The energy of $\nu5/2^+[622]$ steadily becomes smaller with $Z$ increasing both in the
experimental spectra and in the calculated ones.
The lowering of $\nu7/2^+[624]$ with $Z$ increasing is also seen in the experimental
spectra but it is not so striking from the calculation.
In each of these isotones a low-lying state $\nu7/2^-[743]$ is predicted which is not observed.
We note that this prediction is consistent with Ref.~\cite{Parkhomenko2005_APPB36-3115}.

Low-lying one-quasineutron levels for $N=147$ isotones $^{243}$Cm~\cite{Braid1971_PRC4-247}, $^{245}$Cf~\cite{Hessberger2004_EPJA22-417},
and $^{247}$Fm~\cite{Hessberger2006_EPJA30-561} were identified experimentally.
Their ground states are $\nu5/2^+[622]$,
$\nu1/2^+[631]$, and $\nu7/2^+[624]$ respectively.
However, the ground state for all these isotones is $\nu5/2^+[622]$ from the
PNC-CSM as seen in Fig.~\ref{fig:SPLN147}.
The same is obtained recently by using a two-center shell model~\cite{Adamian2011_PRC84-024324}.
In our calculations, the energies of these three levels are very close to each other,
a small change in the deformation parameters would modify the order of these levels.
This might be one of the reasons for the disagreement in the ground state configuration
between the theory and the experiment.
Similar situation happens in the Bk and Es isotopes, in which the energies of
$\pi 3/2^-[521]$ and $\pi 7/2^+[633]$ are very close to each other and these
two one-quasiproton levels cross each other at $N = 152$.

Low-lying one-quasineutron levels for $N=149$ isotones $^{245}$Cm~\cite{Firestone1999},
$^{247}$Cf~\cite{Akovali2004_NDS102-515}, $^{249}$Fm~\cite{Hessberger2007_EPJD45-33} and
$^{251}$No~\cite{Hessberger2006_EPJA30-561} were identified experimentally.
All their ground states are $\nu7/2^+[624]$.
The comparison with the experiment is shown in Fig.~\ref{fig:SPLN149}.
Nearly all of these 1-quasineutron bandhead energies are well reproduced
by the PNC-CSM calculations.
The energy of $\nu1/2^+[631]$ is much lower in ${}^{251}$No than that in ${}^{245}$Cm,
but in our calculations this level seems almost unchange with the proton number increasing.
The calculated results are very similar with that in  Ref.~\cite{Asai2011_PRC83-014315}.
For each observed level, the energy reaches maximum in $^{247}$Cf,
which can not be reproduced by our calculations.

Many theoretical models predict that the first excited state in $N = 151$ isotones
should be $\nu7/2^+[624]$ (see, e.g., Ref.~\cite{Parkhomenko2005_APPB36-3115}).
This is not consistent with experimental results, i.e., the
first excited state in $N=151$ isotones is $\nu5/2^+[622]$.
The low excitation energy of the $\nu5/2^+[622]$ states in the $N = 151$
isotones have been interpreted as a consequence of the presence of a
low-lying $K^\pi = 2^-$ octupole phonon state~\cite{Yates1975_PRC12-442}.
Noted that in Ref.~\cite{Afanasjev2003_PRC67-024309}, by using the cranked relativistic
Hartree-Bogoliubov theory, the level sequence in $N = 151$ isotones is consistent with
the experiment, but it can not reproduce the level sequence in $N = 149$ isotones.
It can be seen in Fig.~\ref{fig:SPLN151} that, the level sequence in most nuclei
is consistent with the experimental data,
with the obviously exception of the $\nu5/2^+[622]$ orbital.
The level $\nu7/2^+[624]$ observed in ${}^{253}$No at 355~keV is from Ref.~\cite{Reiter2005_PRL95-032501}.
A recent experiment~\cite{Herzberg2009_EPJA42-333} observed a very similar level
scheme which was, however, assigned as the $\nu7/2^-[734]$ configuration.
This will be discussed in Sec.~\ref{Sec:results}.

The experimental and calculated bandhead energies of low-lying one-quasineutron
bands for the $N=153$ isotones are compared in Fig.~\ref{fig:SPLN153}.
The one-quasineutron levels for $^{249}$Cm~\cite{Ishii2008_PRC78-054309},
$^{251}$Cf~\cite{Ahmad2005_PRC72-054308}, $^{253}$Fm~\cite{Asai2005_PRL95-102502} and
$^{255}$No~\cite{Hessberger2006_EPJA29-165} were identified experimentally.
The ground states for all these isotones are $\nu1/2^+[620]$ in our calculations,
which is consistent with the data.
But the level order of $\nu7/2^+[613]$ and $\nu3/2^+[622]$ from our calculation
is inversed according to the data.
This also happens in the Ref.~\cite{Parkhomenko2005_APPB36-3115}.
Another problem is that $\nu11/2^-[725]$ states in all the isotones from our
calculation are systematically higher than the experiment values.
In each of these isotones a low-lying state $\nu7/2^-[743]$ is predicted which is not observed.

Low-lying one-quasiproton levels for Bk ($Z=97$) isotopes $^{243,245}$Bk~\cite{Yuichi1989_NPA500-90},
$^{247}$Bk~\cite{Ahmad1979_PRC20-290}, $^{249}$Bk~\cite{Ahmad2005_PRC71-054305}
and $^{251}$Bk~\cite{Tuli2006_NDS107-1347} were identified experimentally.
The comparison between the data and our calculation is shown in Fig.~\ref{fig:SPLBk}.
The two levels $\pi3/2^-[521]$ and $\pi7/2^+[633]$ are very close to each other,
usually the energy difference is less than 50~keV.
The ground states are $\pi3/2^-[521]$ for all Bk isotopes except $^{249}$Bk.
In our calculations the ground states are all $\pi3/2^-[521]$.
The deviation in $^{249}$Bk may be due to the staggering of the deformation.
Nearly all the calculated 1-quasiproton energies in the Bk isotopes are a
little larger than the experimental values.

Low-lying one-quasineutron levels for Es ($Z=99$) isotopes $^{245, 247, 249, 251, 253}$Es~
\cite{Ahmad1978_PRC17-2163, Hessberger2005_EPJA26-233} were identified experimentally.
The comparison between the data and our calculation is shown in Fig.~\ref{fig:SPLEs}.
The ground states are $\pi7/2^+[633]$ for all Es isotopes except $^{251}$Es.
In our calculations the ground states are all $\pi7/2^+[633]$.
The deviation in $^{251}$Es may be also due to the staggering of the deformation.
Our calculations show that the bandhead energies of the negative parity states
are all very small (less than 400~keV), which is consistent with the experiment.

With the proton number increasing, the data become less and less.
Fig.~\ref{fig:SPLMdLr} shows the comparison between the experimental values
and our calculation for the Md and Lr isotopes.
The data are available only for $^{251}$Md~\cite{Chatillon2006_EPJA30-397},
$^{255}$Md~\cite{Ahmad2000_PRC61-044301} and $^{251}$Lr~\cite{Chatillon2006_EPJA30-397}
and they are reproduced by the theory quite well.

From the above discussions, we see that the new Nilsson parameters
can describe satisfactorily the 1-qp spectra of nuclei with $Z \approx 100$.
But there are still some discrepancies.
According to our experience, it is quite difficult to improve this
situation in the framework of the Nilsson model.
One way out might be to use the Woods-Saxon potential instead of the Nilsson potential.

\subsection{Even-even nuclei}

\begin{figure*}[!]
\includegraphics[scale=0.7]{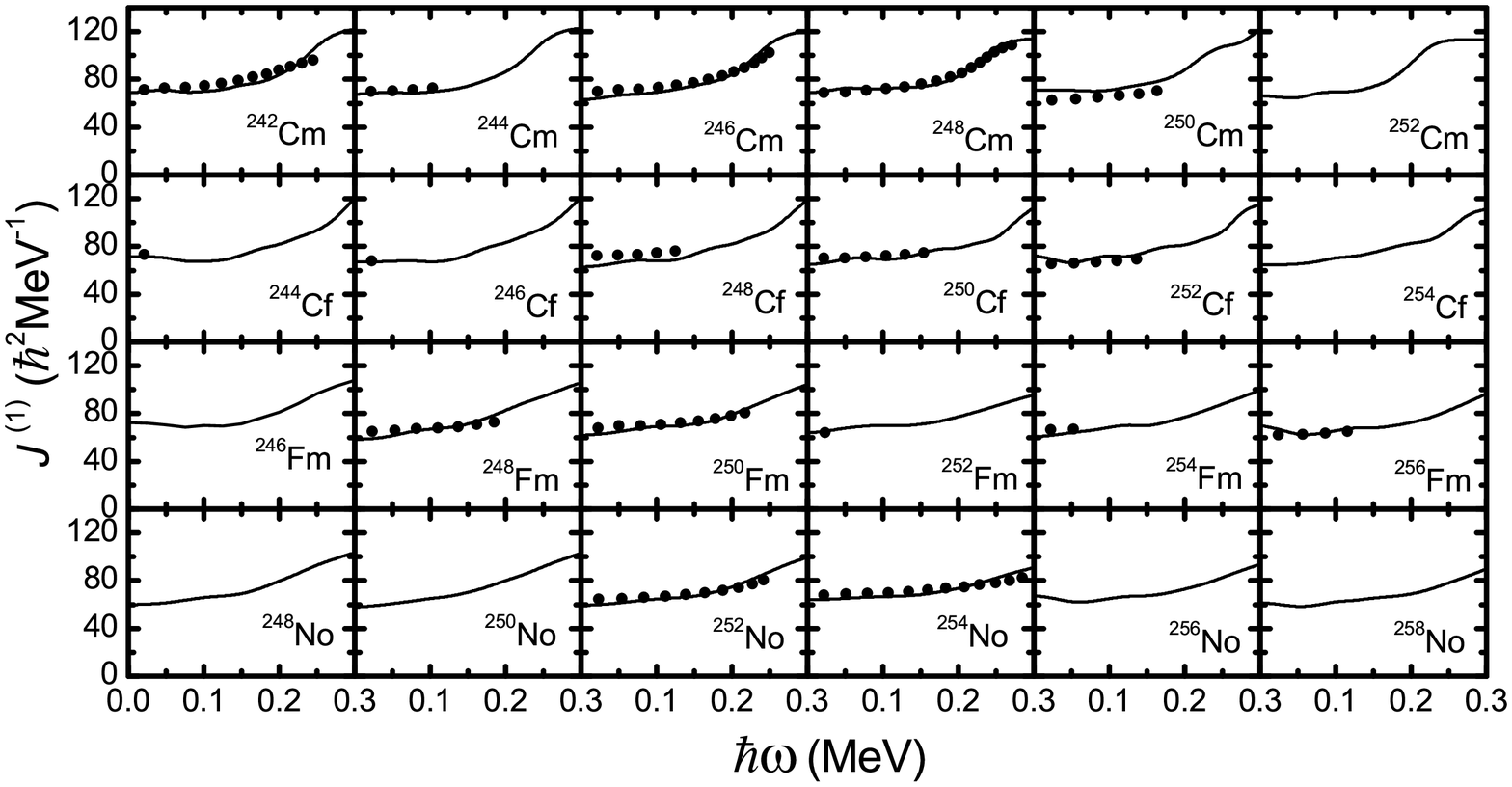}
\caption{\label{fig:MOIeveneven} The experimental (solid circles) and calculated (solid lines)
MOI's $J^{(1)}$ for the GSB's in Cm, Cf, Fm, and No isotopes from $N=146$ to $N=156$.
The data are taken from Ref.~\cite{Herzberg2008_PPNP61-674} and references therein.
The most recent data of $^{248, 250, 252}$Cf are taken from ~\cite{Takahashi2010_PRC81-057303}.
The effective pairing interaction
strengths for both protons and neutrons for all these even-even nuclei are,
$G_n=0.30$~MeV, $G_{2n}=0.020$~MeV, $G_p=0.40$~MeV, and $G_{2p}=0.035$~MeV.}
\end{figure*}

The experimental kinematic MOI's for each band are extracted by
\begin{equation}
 \frac{J^{(1)}(I)}{\hbar^2} = \frac{2I+1}{E_{\gamma}(I+1\rightarrow I-1)} \ ,
\end{equation}
separately for each signature sequence within a rotational band.
The relation between the rotational frequency
$\omega$ and the angular momentum $I$ is
\begin{equation}
 \hbar\omega(I) = \frac{E_{\gamma}(I+1\rightarrow I-1)}{I_{x}(I+1)-I_{x}(I-1)} \ ,
\end{equation}
where $I_{x}(I)=\sqrt{(I+1/2)^{2}-K^{2}}$, $K$ is the projection of nuclear total angular
momentum along the symmetry $z$ axis of an axially symmetric nuclei.

Figure~\ref{fig:MOIeveneven} shows the experimental and calculated MOI's of the GSB's
in even-even Cm, Cf, Fm, and No isotopes.
The experimental (calculated) MOI's are denoted by solid circles (solid lines).
The experimental MOI's of all these GSB's are well reproduced by the PNC-CSM calculations.
For some nuclei, there is no data (i.e., $^{252}$Cm), we only show the PNC-CSM results.
It can be seen from Fig.~\ref{fig:MOIeveneven} that with increasing proton number $Z$
in any isotonic chain, the upbendings in each nuclei become less pronounced.
It is well known that the backbending is caused by the crossing of the GSB with
a pair-broken band based on high-$j$ intruder orbitals~\cite{Stephens1975_RMP47-43},
in this mass region the $\pi i_{13/2}$ and $\nu j_{15/2}$ orbitals.
But in several nuclei, there is no evidence for a $\nu j_{15/2}$ alignment.
It has been pointed out in Ref.~\cite{Zhang2011_PRC83-011304R} that, for the nuclei with
$N\approx150$, among the neutron orbitals of $j_{15/2}$ parentage, only the
high-$\Omega$ (deformation aligned) $\nu 7/2^-[743]$ and $\nu
9/2^-[734]$ are close to the Fermi surface.
The diagonal parts in Eq.~(\ref{eq:jx}) of these two orbitals contribute no
alignment to the upbending, only the off-diagonal parts in Eq.~(\ref{eq:jx})
contribute a little if the neutron $j_{15/2}$ orbital is not blocked.

\begin{figure}
\includegraphics[scale=0.35]{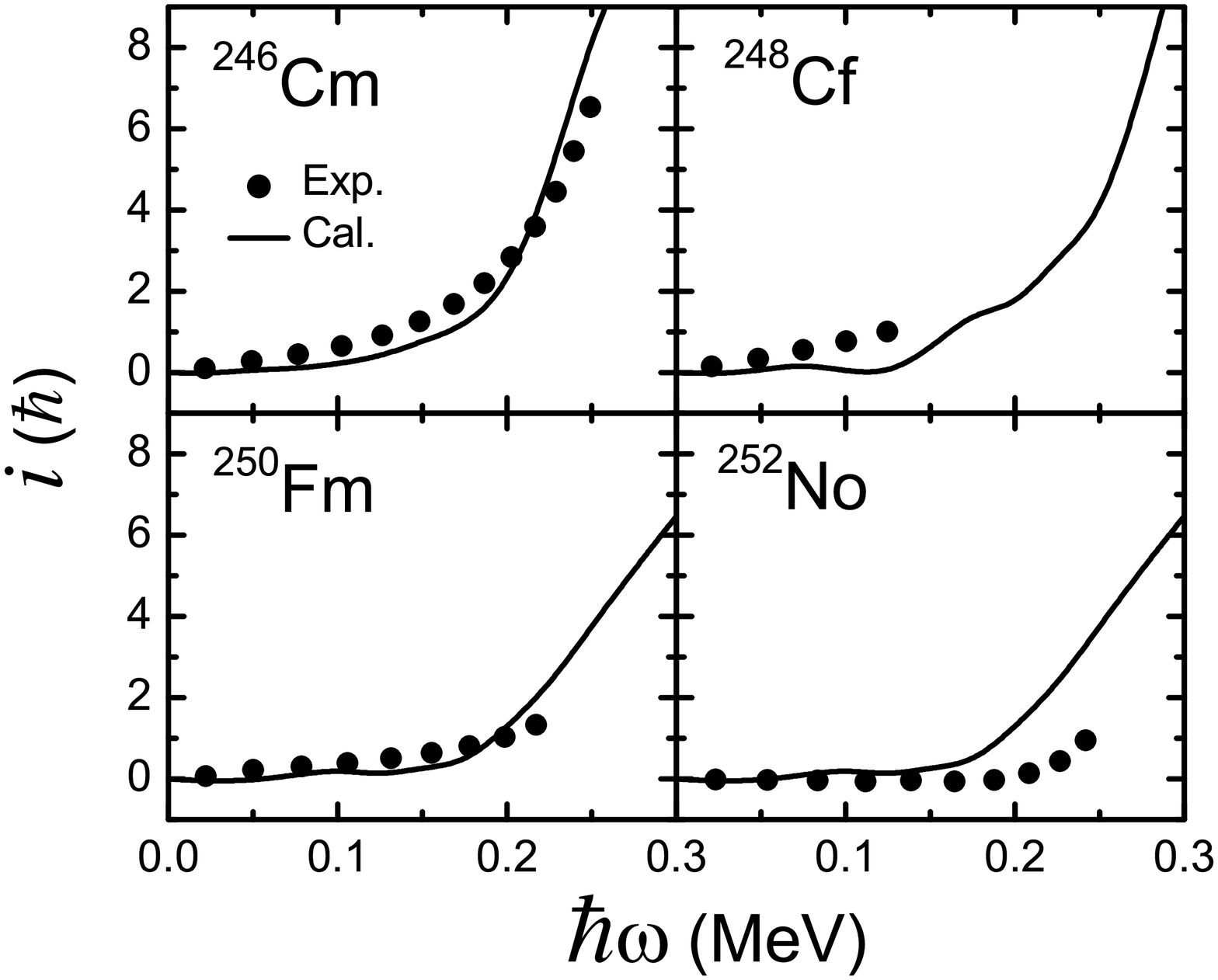}
\caption{\label{fig:Alignment} The experimental (solid circles) and
calculated (solid lines) alignment  $i= \langle
J_x \rangle -\omega J_0 -\omega ^ 3 J_1$ for the GSB's
in $^{246}$Cm, $^{248}$Cf, $^{250}$Fm, and $^{252}$No ($N=150$ isotones).  The Harris parameters
$J_0 = 65\ \hbar^2$MeV$^{-1}$ and $J_1 = 200\ \hbar^4$MeV$^{-3}$. }
\end{figure}

To see the upbending in these even-even nuclei more clearly,
we show the experimental (solid circles) and calculated (solid lines) alignment  $i$ for the GSB's
in $N=150$ isotones in Fig.~\ref{fig:Alignment}.
For other isotones the results are very similar, so we do not shown them here.
It can be seen that the upbending frequencies of these GSB's are about $0.20 \sim 0.25$~MeV.

\begin{figure*}
\includegraphics[scale=0.7]{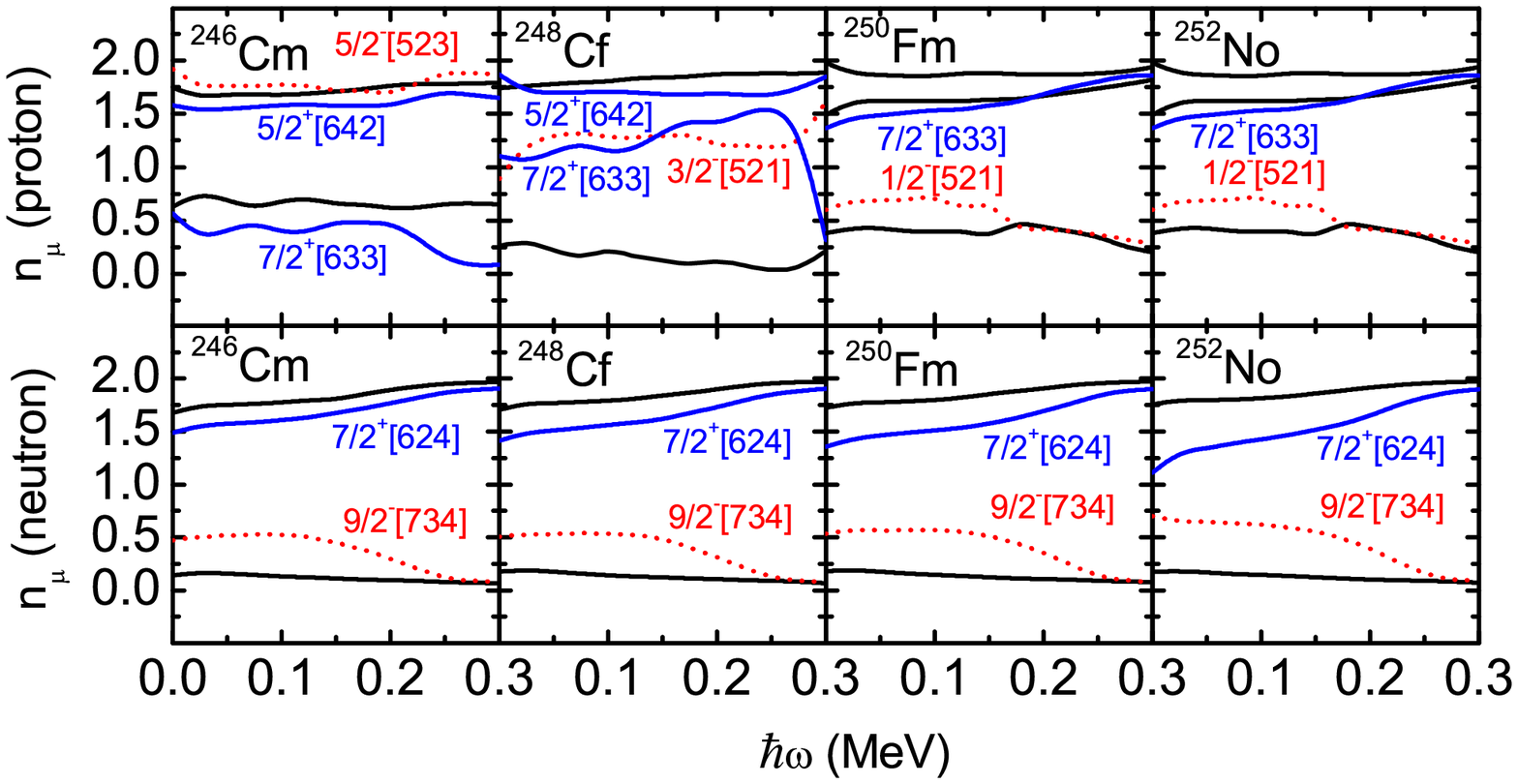}
\caption{\label{fig:Occupation}(Color online) Occupation probability $n_\mu$ of each orbital
$\mu$ (including both $\alpha=\pm1/2$) near the Fermi surface
for the GSB's in $^{246}$Cm, $^{248}$Cf, $^{250}$Fm, and $^{252}$No ($N=150$ isotones).
The top and bottom rows are for protons and neutrons respectively.
The positive (negative) parity levels are denoted by blue solid (red dotted) lines.
The Nilsson levels far above the Fermi surface
($n_{\mu}\sim0$) and far below ($n_{\mu}\sim2$) are not shown. }
\end{figure*}

One of the advantages of the PNC method is that the total particle number
$N = \sum_{\mu}n_\mu$ is exactly conserved, whereas the occupation probability
$n_\mu$ for each orbital varies with rotational frequency $\hbar\omega$.
By examining the $\omega$-dependence of the orbitals close to the Fermi surface, one
can learn more about how the Nilsson levels evolve with rotation and
get some insights on the upbending mechanism.
Fig.~\ref{fig:Occupation} shows the occupation probability $n_\mu$
of each orbital $\mu$ near the Fermi surface for the GSB's in the $N=150$ isotones.
The top and bottom rows are for the protons and neutrons, respectively.
The positive (negative) parity levels are denoted by blue solid (red dotted) lines.
The orbitals in which $n_\mu$ do not change much (i.e., contribute little to the upbending)
are denoted by black lines.
The Nilsson levels far above the Fermi surface ($n_{\mu}\sim0$)
and far below ($n_{\mu}\sim2$) are not shown.
We can see from Fig.~\ref{fig:Occupation} that the occupation probability of
$\pi7/2^+[633]$  ($\pi i_{13/2}$) drops down gradually from 0.5 to nearly zero with
the cranking frequency  $\hbar\omega$ increasing from about $0.20$~MeV to $0.30$~MeV,
while the occupation probabilities of some other orbitals slightly increase.
This can be understood from the cranked Nilsson levels shown in Fig.~\ref{fig:250FmNilsson}.
The $\pi7/2^+[633]$ is slightly above the Fermi surface at $\hbar\omega=0$.
Due to the pairing correlations, this orbital is partly occupied.
With increasing $\hbar\omega$, this orbital leaves farther above the Fermi surface.
So after the band-crossing frequency, the occupation probability of this
orbital becomes smaller with increasing $\hbar\omega$.
Meanwhile, the occupation probabilities of those orbitals which approach near
to the Fermi surface become larger with increasing $\hbar\omega$.
This phenomenon is even more clear in $^{248}$Cf, but the band-crossing occurs
at $\hbar\omega_c\sim0.25$~MeV, a little larger than that of $^{246}$Cm.
So the band-crossings in both cases are mainly caused by the $\pi i_{13/2}$ orbitals.
For $^{250}$Fm and $^{252}$No, the occupation probabilities of $\pi i_{13/2}$ orbitals
increase slowly with the cranking frequency $\hbar\omega$ increasing from about
$0.20$~MeV to $0.30$~MeV.
So there is no sharp bandcrossing from the proton orbitals.
Now we focus on the occupation probability $n_\mu$ of the neutron orbitals (bottom row).
The four figures in the bottom row of Fig.~\ref{fig:Occupation} show a very similar pattern.
It can be seen that, with $\hbar\omega$ increasing, the $n_\mu$ of $\nu 7/2^+[624]$
orbitals increase slowly and that of the high-$\Omega$ (deformation aligned)
$\nu 9/2^-[734]$ orbitals ($j_{15/2}$) decrease slowly.
Thus only a small contribution is expected from neutrons to the
upbendings for the GSB's in the $N=150$ isotopes.
The bandcrossing frequencies for neutrons are about $0.20 \sim 0.25$~MeV,
very close to the proton bandcrossing frequencies.
So neutrons and protons from the high-$j$ orbits compete strongly in rotation-alignment
as pointed out in Ref.~\cite{Al-Khudair2009_PRC79-034320}.

\begin{figure*}
\includegraphics[scale=0.7]{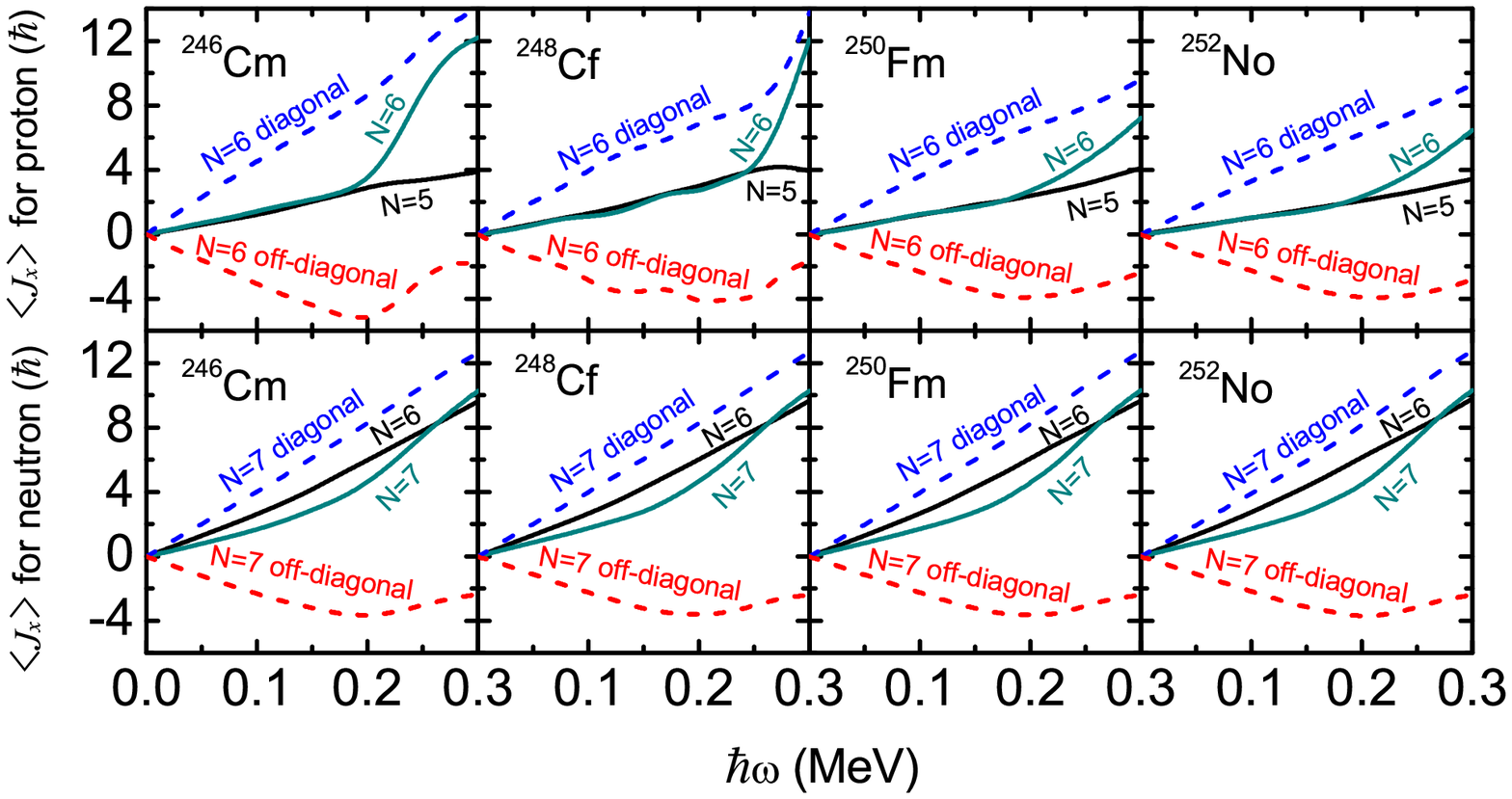}
\caption{\label{fig:Jxshell}(Color online) Contribution of each proton (top row)
and neutron (bottom) major shell to the angular momentum alignment $\langle J_x\rangle$
for the GSB's in $^{246}$Cm, $^{248}$Cf, $^{250}$Fm, and $^{252}$No ($N=150$ isotones).
The diagonal $\sum_{\mu} j_x(\mu)$ and off-diagonal parts $\sum_{\mu<\nu} j_x(\mu\nu)$
in Eq.~(\ref{eq:jx}) from the proton $N=6$ and neutron $N=7$ shells are shown by dashed lines.
}
\end{figure*}

The contribution of each proton (top row) and neutron (bottom row) major shell to the
angular momentum alignment $\langle J_x\rangle$ for the GSB's in
the $N=150$ isotones is shown in Fig.~\ref{fig:Jxshell}.
The diagonal $\sum_{\mu} j_x(\mu)$ and off-diagonal parts $\sum_{\mu<\nu} j_x(\mu\nu)$
in Eq.~(\protect\ref{eq:jx}) from the proton $N=6$ and the neutron $N=7$
shells are shown by dashed lines.
Note that in this figure, the smoothly increasing part of the alignment
represented by the Harris formula ($\omega J_0 +\omega ^ 3 J_1$) is not
subtracted  (cf. the caption of Fig.~\ref{fig:Alignment}).
It can be seen clearly that the upbendings for the GSB's in $^{246}$Cm at
$\hbar\omega_c\sim$ 0.20~MeV and in $^{248}$Cf at $\hbar\omega_c\sim$ 0.25~MeV
mainly come from the contribution of the proton $N=6$ shell.
Furthermore, the upbending for the GSB in $^{246}$Cm is mainly from the off-diagonal part of
the proton $N=6$ shell, while both the diagonal and off-diagonal
parts of the proton $N=6$ shell contribute to the upbending for the GSB in $^{248}$Cf.
The off-diagonal part of the neutron $N=7$ shell only contributes a little to the upbending.
It is very different from $^{250}$Fm and $^{252}$No.
For these two nuclei, the contribution to the upbending from the off-diagonal parts of
the proton $N=6$ shell and the off-diagonal part of the neutron $N=7$ shell is nearly the same.
This is because that, with the proton number $Z$ increasing, the Fermi surface
leaves further and further from the $\pi7/2^+[633]$ orbital.
In this case, the high-$j$ but high-$\Omega$ orbital (deformation aligned)
$\pi9/2^+[624]$ becomes close to the Fermi surface.

\begin{figure*}
\includegraphics[scale=0.7]{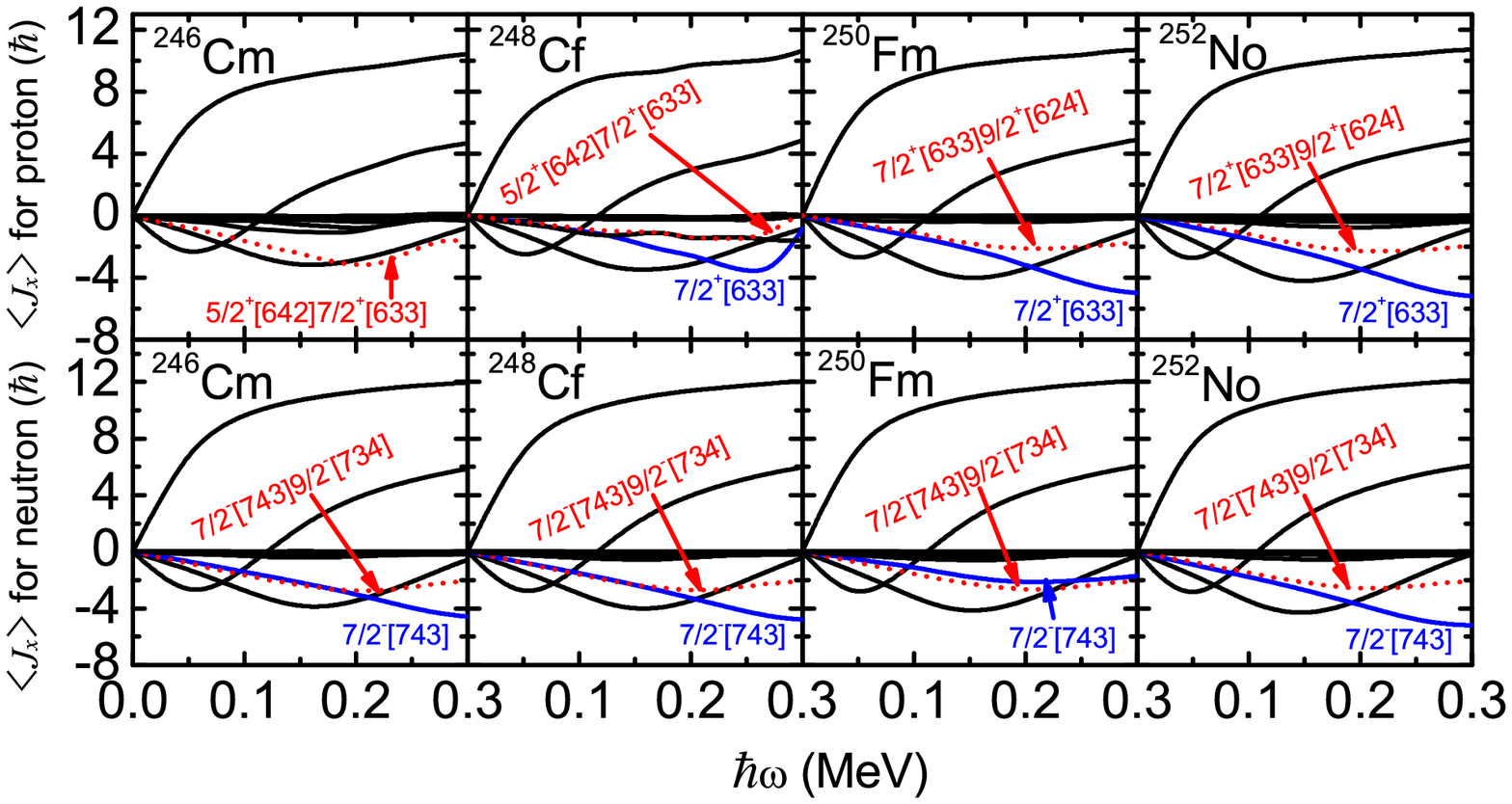}
\caption{\label{fig:Jxorbital}(Color online) Contribution of each proton orbital in the $N=6$
major shell (top row) and each neutron orbital in the $N=7$
major shell (bottom row) to the angular momentum alignments
$\langle J_x\rangle$ for the GSB's in $^{246}$Cm, $^{248}$Cf, $^{250}$Fm,
and $^{252}$No ($N=150$ isotones).
The important diagonal (off-diagonal) part $j_x(\mu)$ [$j_x(\mu\nu)$]
in Eq.~(\ref{eq:jx}) is denoted by blue solid (red dotted) lines.
The orbitals that have no contribution to the upbending
(some of these orbitals contribute to the steady increase of the alignment)
are denoted by black lines.}
\end{figure*}

In order to have a  more clear understanding of the upbending mechanism, the contribution  of intruder
proton orbitals $i_{13/2}$ (top row) and intruder neutron orbitals $j_{15/2}$ (bottom row)
to the angular momentum alignments $\langle J_x\rangle$ is presented in Fig.~\ref{fig:Jxorbital}.
The important diagonal (off-diagonal) part $j_x(\mu)$ [$j_x(\mu\nu)$] in
Eq.~(\protect\ref{eq:jx}) is denoted by blue solid (red dotted) lines.
The orbitals that have no contribution to the upbending
(some of these orbitals contribute to the steady increase of the alignment) are denoted by black lines.
Near the proton Fermi surfaces of $^{246}$Cm and $^{248}$Cf,
the proton $i_{13/2}$ orbitals are $\pi 5/2^+[642]$ and $\pi 7/2^+[633]$.
Other orbitals of $\pi i_{13/2}$ parentage are either fully occupied or fully empty
(cf. Fig.~\ref{fig:Occupation}) and have no contribution to the upbending
(only contribute to the steady increase of the alignment).
For $^{246}$Cm, the PNC calculation shows that after the upbending  ($\hbar \omega \geq$ 0.20 MeV)
the off-diagonal part $j_x\left(\pi 5/2^+[642] \pi 7/2^+[633]\right)$ changes a lot.
The alignment gain after the upbending mainly comes from this interference term.
For $^{248}$Cf, the main contribution to the alignment gain after
the upbending comes from the diagonal part $j_x\left(\pi 7/2^+[633]\right)$
and the off-diagonal part $j_x\left(\pi 5/2^+[642] \pi 7/2^+[633]\right)$.
Again this tells us that the upbending in both cases is mainly caused by the $\pi i_{13/2}$ orbitals.
As to $^{250}$Fm and $^{252}$No, only the off-diagonal part
$j_x\left(\pi 7/2^+[633] \pi 9/2^+[624] \right)$ contributes a little to the upbending.
The absence of the alignment of $j_{15/2}$ neutrons in nuclei in this mass region
can be understood from the contribution of the intruder neutron orbitals ($N=7$) to
$\langle J_x \rangle$. For the nuclei with $N \approx 150$, among
the neutron orbitals of $j_{15/2}$ parentage, only the
high-$\Omega$ (deformation aligned) $\nu 7/2^-[743]$ and $\nu
9/2^-[734]$ are close to the Fermi surface.
The diagonal parts of these two orbitals contribute no alignment to the upbending,
only the interference terms $j_x\left(\nu 7/2^-[743] \pi 9/2^-[734]\right)$
contribute a little to the alignment.
So we can see the strong competition in rotation-alignment between the high-$j$ protons and neutrons
in $^{250}$Fm and $^{252}$No, which is consistent with the result in Ref.~\cite{Al-Khudair2009_PRC79-034320}.

\subsection{Odd-$A$ nuclei}

It is well known that there exist large fluctuations in the experimental
odd-even differences in MOI's $\delta J / J$.
If a high-$j$ intruder orbital near the Fermi surface is blocked
$\delta J / J$ is quite large (sometime larger than 100\%).
It is very hard to reproduce the large fluctuations with the conventional 
BCS method which predicts that $\delta J / J$ is about 
$15\%$~\cite{Bohr1975_Nucl_Structure}.
One of the advantages of the PNC-CSM is that the Pauli blocking 
effects are treated exactly. So the odd-even differences in 
the MOI's $\delta J / J$ can be reproduced quite well
and this has been shown for rare-earth-nuclei~\cite{Zeng1994_PRC50-746}.
There are two high-$j$ orbitals involved in the present calculations,
namely, $\pi 7/2^+[633]$ ($\pi i_{13/2}$) and $\nu 9/2^-[734]$ ($\nu j_{15/2}$).
To show the blocking effects we study the following four nuclei and compare 
the calculated odd-even differences in MOI's with the data,
\begin{eqnarray}
\frac{\delta J}{J} = \frac{J(^{249}{\rm Bk} \, \pi 7/2^+[633]) - J(^{248}{\rm Cm \, GSB})}
     {J(^{248}{\rm Cm \, GSB})}
     \approx 54\% \, ({\rm Exp.})\ , \quad 56\% \, ({\rm Cal.}) \ , \nonumber\\
\frac{\delta J}{J} = \frac{J(^{253}{\rm No} \, \nu 9/2^-[734]) - J(^{252}{\rm No \, GSB})}
     {J(^{252}{\rm No \, GSB})}
     \approx 41\% \, ({\rm Exp.})\ , \quad 46\% \, ({\rm Cal.}) \ . \nonumber
\end{eqnarray}
It can be seen that the experimentally observed large odd-even difference 
in MOI's induced by the high-$j$ intruder orbitals can be reproduced quite well.

\begin{figure*}
\includegraphics[scale=0.6]{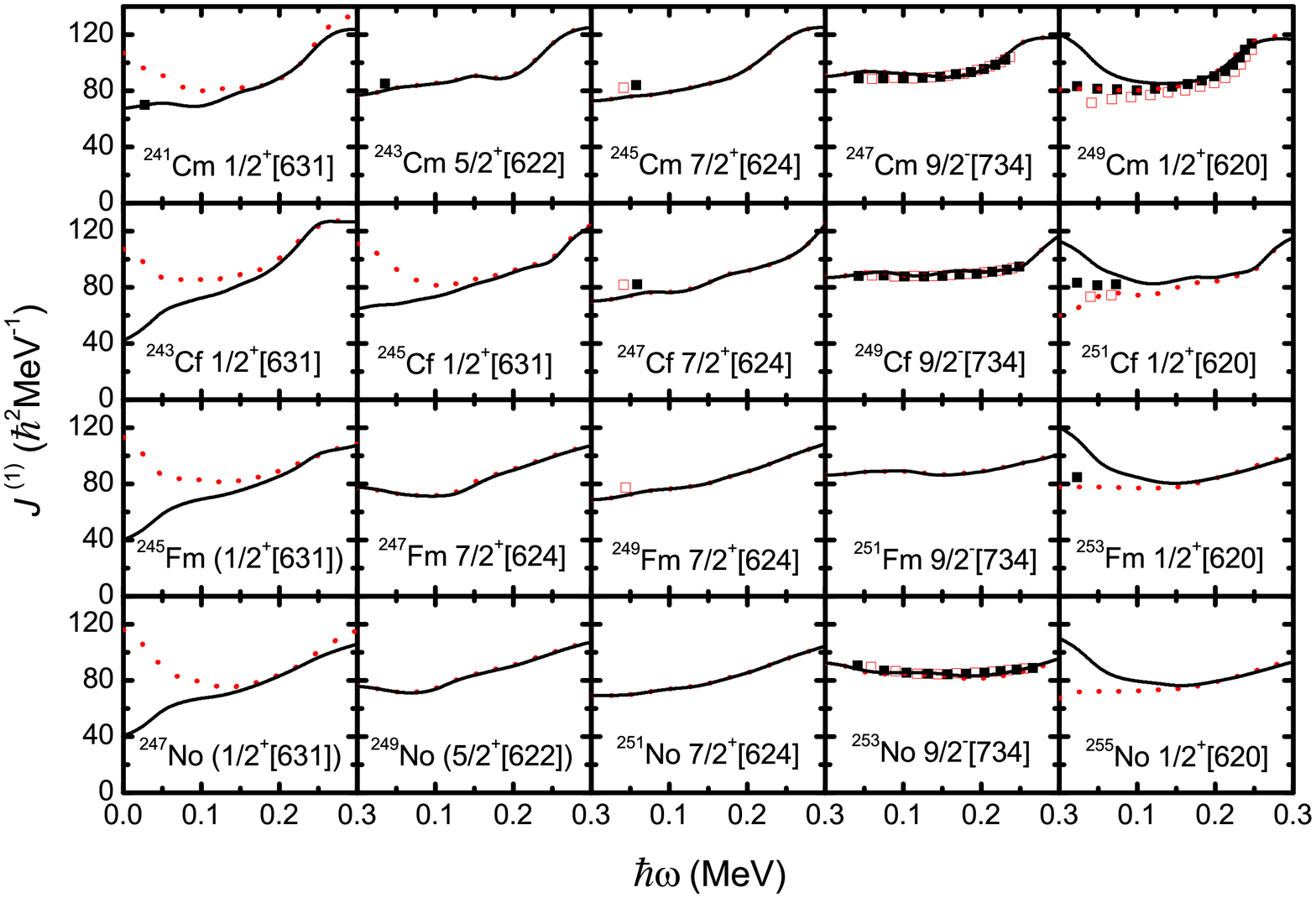}
\caption{\label{fig:MOIoddNg}(Color online) The experimental and calculated MOI's $J^{(1)}$
of the GSB's for odd-$N$ Cm, Cf, Fm, and No isotopes.
The data are taken from Ref.~\cite{Herzberg2008_PPNP61-674} and references therein.
The most recent data of $^{247, 249}$Cm and $^{249}$Cf are taken from ~\cite{Tandel2010_PRC82-041301R}.
The experimental MOI's are denoted by full square
(signature $\alpha=+1/2$) and open square (signature $\alpha=-1/2$), respectively.
The calculated MOI's by the PNC method are denoted by solid lines
(signature $\alpha=+1/2$) and dotted lines (signature $\alpha=-1/2$), respectively.
The GSB's in the bracket denote the PNC-CSM calculated results which have not been observed.
The effective pairing interaction
strengths for both protons and neutrons for all these odd-$N$ nuclei are
$G_n=0.25$~MeV, $G_{2n}=0.015$~MeV, $G_p=0.40$~MeV, and $G_{2p}=0.035$~MeV.}
\end{figure*}

Figure~\ref{fig:MOIoddNg} shows the experimental and calculated MOI's of the GSB's
for odd-$N$ Cm, Cf, Fm, and No isotopes ($N = 145 - 153$).
The experimental MOI's are denoted by solid squares (signature $\alpha=+1/2$)
and open squares (signature $\alpha=-1/2$), respectively.
The calculated MOI's are denoted by solid lines (signature $\alpha=+1/2$)
and dotted lines (signature $\alpha=-1/2$), respectively.
The experimental MOI's of all these 1-qp bands are well reproduced by the PNC-CSM
calculations, which in turn strongly support the configuration assignments for them.
The GSB's in the bracket denote the PNC-CSM calculated results which have not been observed.
We should note that the pairing strength used in Ref.~\cite{Zhang2011_PRC83-011304R}
is a little different with what we used now.
This is because the pairing strengths are considered as an average for all of these odd-$N$ nuclei.

\begin{figure*}
\includegraphics[scale=0.5]{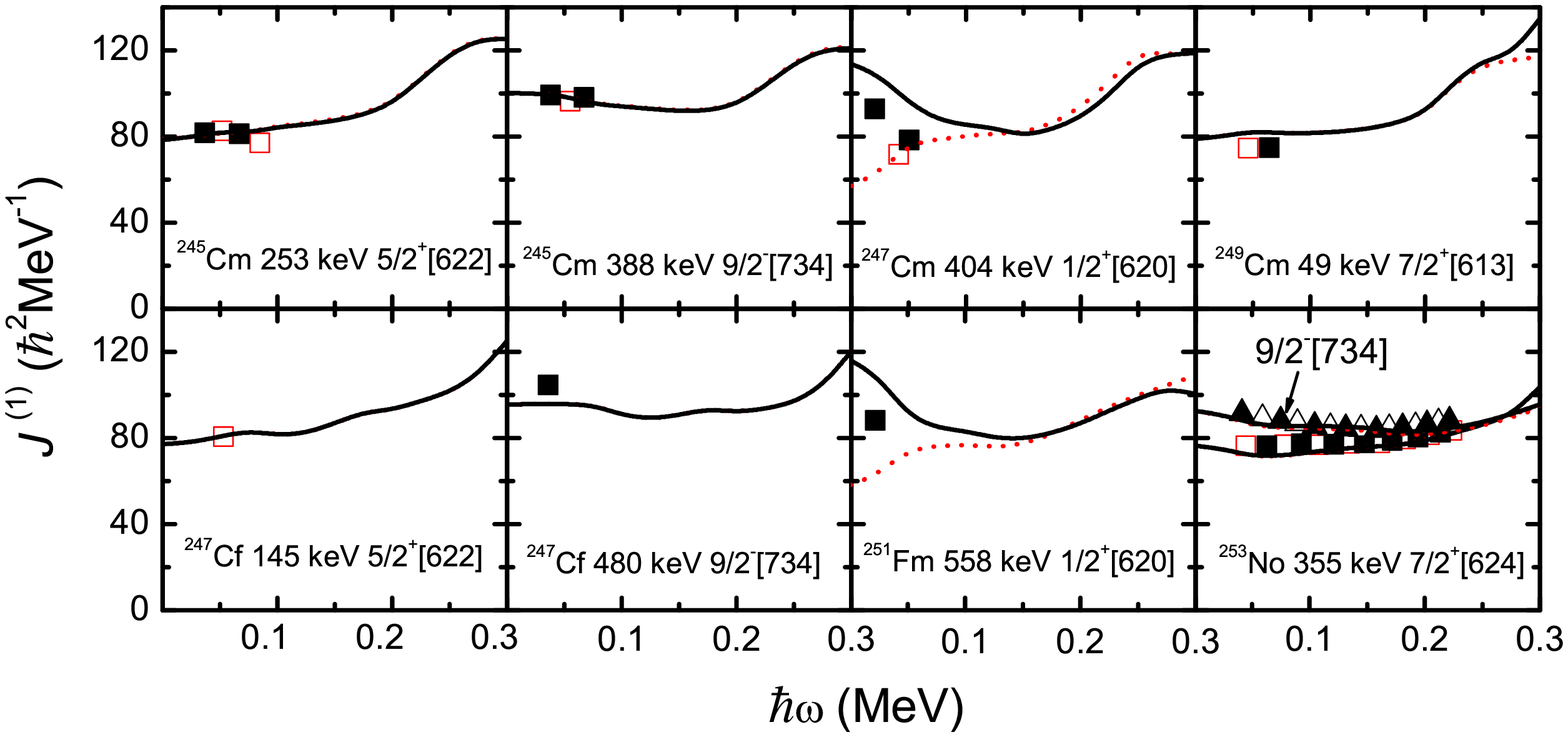}
\caption{\label{fig:MOIoddNE}(Color online) The same as Fig.~\ref{fig:MOIoddNg},
but for the excited 1-qp bands in odd-$N$ Cm, Cf, and No isotopes.
The data of $^{253}$No are taken from~\cite{Reiter2005_PRL95-032501}.
Our calculated results according to two configuration assignments for this band,
$\nu 9/2^-[734]$ and $\nu 7/2^+[624]$, are shown by squares and triangles, respectively.}
\end{figure*}

In the $N=145$ isotones, our calculations predict a significant signature splitting
at the low rotational frequency ($\hbar \omega < 0.20$~MeV) for the $\nu 1/2^+[631]$ orbital.
In the $N=153$ isotones, the signature splitting of the
$\nu 1/2^+[620]$ orbital is well reproduced by our calculation, too.
It is understandable from the behavior of the cranked Nilsson orbital
$\nu 1/2^+[631]$ and $\nu1/2^+[620]$ in Fig.~\ref{fig:250FmNilsson}.
In Ref.~\cite{Zhang2011_PRC83-011304R}, we have already analyzed the upbending
mechanism for the high-spin rotational GSB's of ${}^{247,249}$Cm and $^{249}$Cf observed in
Ref.~\cite{Tandel2010_PRC82-041301R}.

Figure~\ref{fig:MOIoddNE} shows the results of excited 1-qp bands observed
in the odd-$A$ Cm, Cf, Fm, and No isotopes.
They are all well reproduced by the PNC-CSM calculation.
It is interesting to noted that, in an earlier experiment for $^{253}$No,
rotational band has been established and the configuration was assigned as $\nu 7/2^+[624]$.
In a latter experiment, a similar rotational band has been observed~\cite{Herzberg2009_EPJA42-333},
but the configuration was assigned as $\nu 9/2^-[734]$.
It can be seen that, the experimental MOI's extracted from~\cite{Reiter2005_PRL95-032501}
using these two configurations can be reproduced by the PNC calculations.
So from our calculation, it can not be distinguished which configuration assignment is correct.

\begin{figure*}
\includegraphics[scale=0.6]{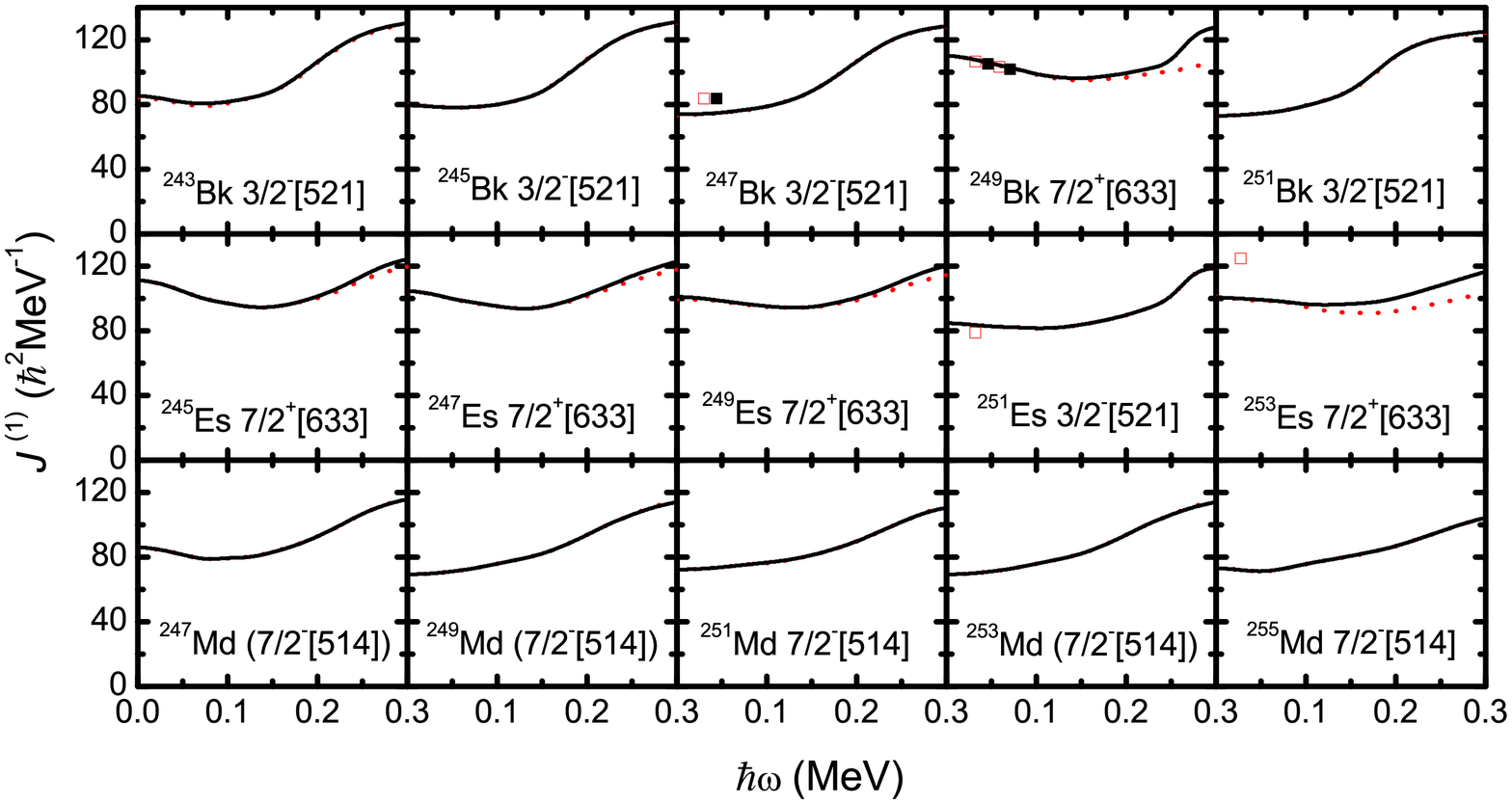}
\caption{\label{fig:MOIoddZg}(Color online) The experimental and
calculated MOI's $J^{(1)}$ of the GSB's for odd-$Z$ Bk, Es, and Md isotopes.
The data are taken from Ref.~\cite{Herzberg2008_PPNP61-674} and references therein.
The experimental MOI's are denoted by full square
(signature $\alpha=+1/2$) and open square (signature $\alpha=-1/2$), respectively.
The calculated MOI's by the PNC method are denoted by solid lines
(signature $\alpha=+1/2$) and dotted lines (signature $\alpha=-1/2$), respectively.
The states in the bracket denote that the GSB's have not been observed.
The effective pairing interaction
strengths for both protons and neutrons for all these odd-$N$ nuclei are,
$G_n=0.30$~MeV, $G_{2n}=0.02$~MeV, $G_p=0.25$~MeV, and $G_{2p}=0.01$~MeV.}
\end{figure*}

\begin{figure}
\includegraphics[scale=0.3]{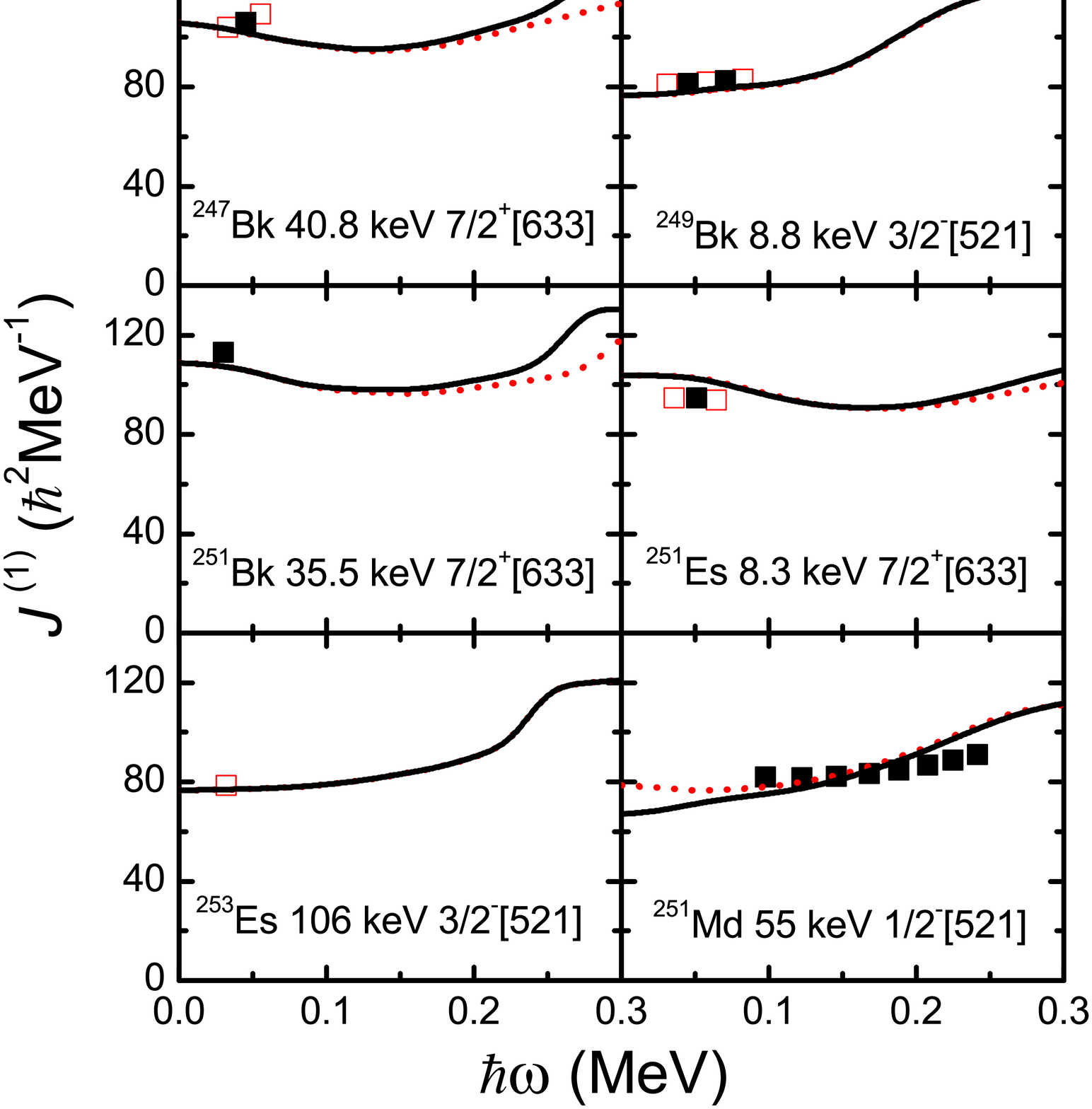}
\caption{\label{fig:MOIoddZE}(Color online) The same as Fig.~\ref{fig:MOIoddZg},
but for the excited 1-qp bands in odd-$Z$ Bk, Es, and Md isotopes.}
\end{figure}

Figure~\ref{fig:MOIoddZg} is the experimental and calculated GSB's of the
odd-$Z$ nuclei, including Bk, Es, and Md isotopes ($N = 146 - 154$). The data for
these odd-$Z$ nuclei are very rare. Most of the data are well reproduced, except
the $\pi 7/2^+[633]$ band in $^{253}$Es.
The MOI of this band seems extremely large, $J^{(1)}$ is larger than 120~$\hbar^2$MeV$^{-1}$.
Our calculations show that the $\pi 7/2^+[633]$
has a small signature splitting at the higher rotational frequency ($\hbar \omega > 0.10$~MeV).
Figure.~\ref{fig:MOIoddZE} shows the results of excited 1-qp bands
observed in the odd-$Z$ Bk, Es, and Md isotopes.

In some nuclei there exists a upbending around $\hbar \omega \approx 0.25$~MeV in the
$\pi 7/2^+[633]$ (signature $\alpha= 1/2$) band according to the PNC calculations
(i.e., the Bk isotopes).
We show the occupation probabilities of the $n_\mu$ of each orbital
$\mu$ near the Fermi surface of the $\pi 7/2^+[633]$ ($\alpha= 1/2$) band in $^{249}$Bk.
The top and bottom rows are for protons and neutrons respectively.
The positive (negative) parity levels are denoted by blue solid (red dotted) lines.
It is clearly to see that the occupation probabilities of $\pi 5/2^+[642]$ and $\pi 7/2^+[633]$
orbitals drop sharply at $\hbar \omega \approx 0.25$~MeV,
while $n_\mu$ of $\pi 3/2^-[521]$ increases sharply.
The occupation probabilities of the neutron orbitals change slowly with increasing $\hbar \omega$,
indicating little contribution to the upbending.
Figure.~\ref{fig:249BkJxShell} shows the contribution of each proton (top row)
and neutron (bottom row) major shell to the angular momentum alignment
$\langle J_x\rangle$ for the $\pi 7/2^+[633]$ ($\alpha= 1/2$) band in $^{249}$Bk.
The diagonal $\sum_{\mu} j_x(\mu)$ and off-diagonal parts
$\sum_{\mu<\nu} j_x(\mu\nu)$ in Eq.~(\ref{eq:jx}) from the proton $N=6$ and neutron $N=7$
shells are shown by dashed lines.
It can be seen from Fig.~\ref{fig:249BkJxShell} that, both the
diagonal and off-diagonal part from proton $N=6$ shell contribution to the upbending.
PNC calculation shows that the upbending comes from the diagonal parts $j_x\left(\pi
5/2^+[642]\right)$ and $j_x\left(\pi 7/2^+[633]\right)$, and the off-diagonal parts
$j_x\left(\pi 5/2^+[642] \pi 7/2^+[633]\right)$ and
$j_x\left(\pi 7/2^+[633] \pi 9/2^+[624] \right)$.

\begin{figure}
\includegraphics[scale=0.23]{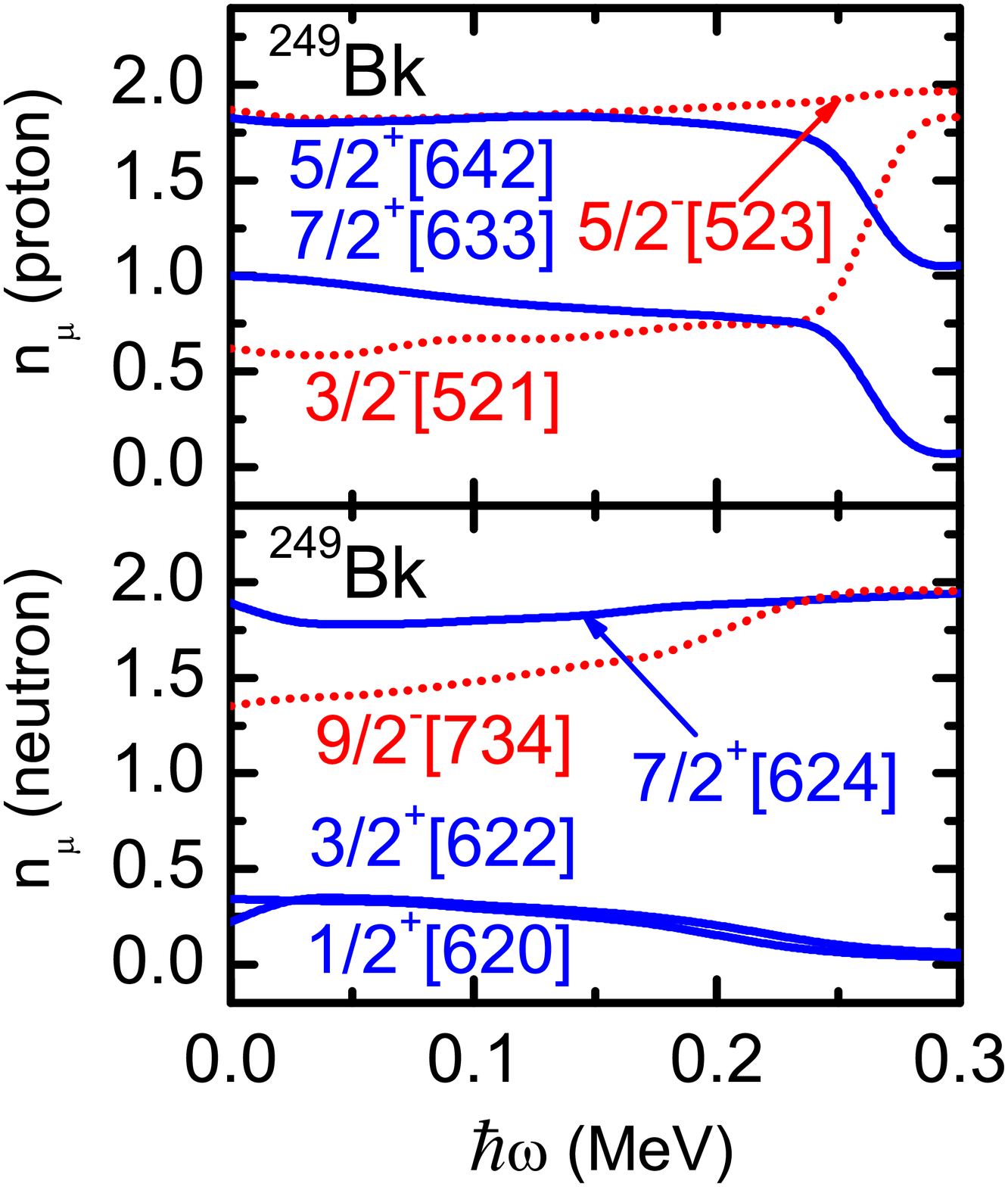}
\caption{\label{fig:249BkOccupation}  Occupation probability $n_\mu$ of each orbital
$\mu$ near the Fermi surface of the $\pi 7/2^+[633]$ ($\alpha= 1/2$) band in $^{249}$Bk.
The top and bottom rows are for protons and neutrons respectively.
The positive (negative) parity levels are denoted by blue solid (red dotted) lines.
The Nilsson levels far above the Fermi surface
($n_{\mu}\sim0$) and far below ($n_{\mu}\sim2$) are not shown. }
\end{figure}

\begin{figure}
\includegraphics[scale=0.23]{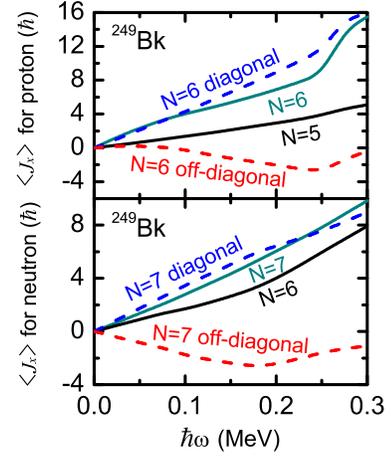}
\caption{\label{fig:249BkJxShell}(Color online) Contribution of each proton (top row)
and neutron (bottom row) major shell to the angular momentum alignment
$\langle J_x\rangle$ for the $\pi 7/2^+[633]$ ($\alpha= 1/2$) band in $^{249}$Bk.
The diagonal $\sum_{\mu} j_x(\mu)$ and
off-diagonal parts $\sum_{\mu<\nu} j_x(\mu\nu)$ in Eq.~(\ref{eq:jx})
from the proton $N=6$ and neutron $N=7$ shells are shown by dashed lines.}
\end{figure}

\subsection{Odd-odd nuclei}

\begin{figure*}
\includegraphics[scale=0.5]{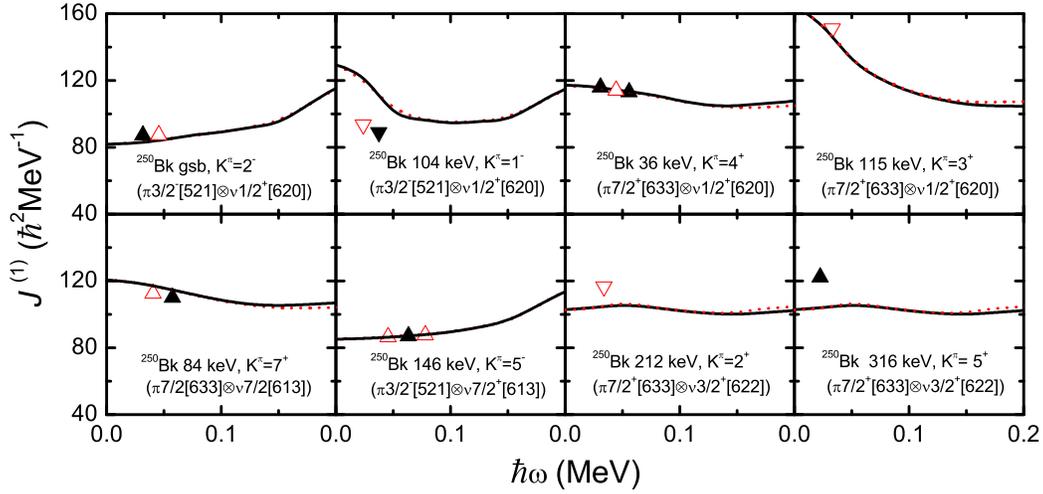}
\caption{\label{fig:250Bk}(Color online) The experimental and calculated MOI's of $^{250}$Bk.
The data are taken from~\cite{Firestone1999, Ahmad2008_PRC77-054302}.
The up (down) triangles denote the experimental value of $K_>$ ($K_<$) bands.
The filled (open) triangles denote the signature $\alpha$ = 0 (1) bands.
The solid (dotted) lines denote the calculated results with signature $\alpha$ = 0 (1) bands.
The effective pairing interaction strengths for both protons and neutrons for all these odd-$N$ nuclei are, $G_n=0.25$~MeV, $G_{2n}=0.015$~MeV, $G_p=0.25$~MeV, and $G_{2p}=0.010$~MeV.}
\end{figure*}

When an unpaired proton and an unpaired neutron in a
deformed odd-odd nucleus are coupled, the projections of their total
angular momentum on the nuclear symmetry axis, $\Omega_p$ and $\Omega_n$, can
produce two states with $K_> =|\Omega_p+ \Omega_n|$ and $K_< =|\Omega_p - \Omega_n|$.
They follow the Gallagher-Moszkowski (GM) coupling rules~\cite{Gallagher1958_PR0111-1282}:
\begin{eqnarray}
 K_> &=& |\Omega_p + \Omega_n|, \ \text{if} \ \Omega_p=\Lambda_p \pm \frac{1}{2} \
                                 \text{and} \ \Omega_n=\Lambda_n \pm \frac{1}{2} \ ,
\nonumber\\
 K_< &=& |\Omega_p - \Omega_n|, \ \text{if} \ \Omega_p=\Lambda_p \pm \frac{1}{2} \
                                 \text{and} \ \Omega_n=\Lambda_n \mp \frac{1}{2} \ .
\nonumber
\end{eqnarray}

The rotational bands in odd-odd nuclei in the transfermium region are very rare.
The most recent experiment is $^{250}$Bk~\cite{Ahmad2008_PRC77-054302}.
In ~\cite{Ahmad2008_PRC77-054302}, the energy splittings between parallel and antiparallel coupled
neutron and proton states were measured for six pairs of states.
Because the residual proton-neutron interaction is not included in our calculations,
the energies of the 2-qp bands in odd-odd
nuclei are the sum of the quasiproton and the quasineutron, that is to say, there is no
energy splitting between states with parallel and antiparallel coupling.

Figure~\ref{fig:250Bk} shows the experimental and calculated MOI's of $^{250}$Bk.
The energy of neutron orbital $\nu1/2^-[750]$ is too high in our calculation,
so we do not take into account the $K^\pi=3^-, 4^-$($\pi7/2^+[633] \otimes \nu1/2^-[750]$)
bands observed in Ref.~\cite{Ahmad2008_PRC77-054302}.
The up (down) triangles denote the experimental value of $K_>$ ($K_<$) bands.
The filled (open) triangles denote the signature $\alpha$ = 0 (1) bands.
The solid (dotted) lines denote the calculated results with signature $\alpha$ = 0 (1) bands.
Most of the data are reproduced quite well.
The calculated MOI's are a little smaller in the $K^\pi = 2^+, 5^+$ bands than the experimental values,
while much larger in the $K^\pi = 1^-$ band.

The MOI's of a $K_>$ and the corresponding $K_<$ bands are usually the same.
But the MOI's of $K^\pi = 3^+$ band are much larger than those of the $K^\pi = 4^+$ band.
This is because for the $K^\pi = 3^+$ band, the neutron component is
$\nu1/2^+[620] (\alpha = +1/2)$, whereas for the $K^\pi = 4^+$ band, the neutron component is
$\nu1/2^+[620] (\alpha = -1/2)$ and the signature splitting of $\nu1/2^+[620]$ is very large.
The calculated results are very close to the data.
Similarly, our calculation predicts that this signature splitting results in
a big differences between the MOI's of the $K^\pi = 1^-$ and $2^-$ bands.
However, this is not the case in the experiment.
It is well known that when one of the nucleons is in an $\Omega=1/2$ orbital, the GM doublet
has $\Delta K = 1$, accordingly the two bands are expected to be Coriolis admixed.
This effect can be very significant in the $K_<$ band, which has been identified in the
odd-odd rare-earth nuclei~\cite{O'neil1972_NPA195-207, Jain1998_RMP70-843}.
The similar MOI's in the $K^\pi = 1^-$ and $2^-$ bands may be from this mixing
and need further exploration both experimentally and theoretically.

\section{Summary}{\label{Sec:summary}}

In summary, the rotational bands in the $A \approx 250$ mass region
are investigated using a cranked shell model (CSM) with the pairing correlations
treated by a particle-number conserving (PNC) method.
In the PNC method for the
pairing correlations, the blocking effects are taken into account exactly.
By fitting the experimental single-particle spectra in
these nuclei, a new set of Nilsson parameters ($\kappa$ and $\mu$) and
deformation parameters ($\varepsilon_2$ and $\varepsilon_4$) are proposed.
The bandhead energies of 1-qp states with energies less than
0.8~MeV are reproduced satisfactorily.
The experimentally observed $\omega$ variations of MOI's for the even-even, odd-$A$, and
odd-odd nuclei are reproduced very well by the PNC-CSM calculations.
By analyzing the $\omega$-dependence of the occupation probability of each
cranked Nilsson orbital near the Fermi surface and the
contributions of valence orbitals in each major shell to the
angular momentum alignment, the upbending
mechanism in this region is understood clearly.
For Cm and Cf isotopes, the upbending in the GSB's is mainly
caused by the intruder proton ($N=6$) $\pi i_{13/2}$ orbitals.
For Fm and No isotopes, neutrons and protons from the high-$j$ orbits
compete strongly in rotation-alignment.
The 2-qp states in the odd-odd nuclei $^{250}$Bk are analyzed in detail.

\begin{acknowledgements}
Helpful discussions with
G. Adamian, N. Antonenko, B. A. Brown, N. V. Giai, R. V. Jolos, J. Meng,
P. Ring, F. Sakata, Y. Sun, and D. Vretenar
are gratefully acknowledged.
This work has been supported by
NSFC (Grant Nos. 10705014, 10875157, 10975100, 10979066, 11175252, and 11120101005),
MOST (973 Project 2007CB815000),
CAS (Grant Nos. KJCX2-EW-N01 and KJCX2-YW-N32),
and FRF for NUAA (Grant No. NS2010157).
The computation of this work
was supported by Supercomputing Center, CNIC of CAS.
\end{acknowledgements}



\bibliographystyle{apsrev4-1}

%

\end{document}